Editors' Suggestion | Featured in Physics

# GW190521: A Binary Black Hole Merger with a Total Mass of 150 $M_\odot$

R. Abbott et al.*
(LIGO Scientific Collaboration and Virgo Collaboration)

On May 21, 2019 at 03:02:29 UTC Advanced LIGO and Advanced Virgo observed a short duration gravitational-wave signal, GW190521, with a three-detector network signal-to-noise ratio of 14.7, and an estimated false-alarm rate of 1 in 4900 yr using a search sensitive to generic transients. If GW190521 is from a quasicircular binary inspiral, then the detected signal is consistent with the merger of two black holes with masses of $85^{+21}_{-14}\ M_\odot$ and $66^{+17}_{-18}\ M_\odot$ (90% credible intervals). We infer that the primary black hole mass lies within the gap produced by (pulsational) pair-instability supernova processes, with only a 0.32% probability of being below 65 $M_\odot$. We calculate the mass of the remnant to be $142^{+28}_{-16}\ M_\odot$, which can be considered an intermediate mass black hole (IMBH). The luminosity distance of the source is $5.3^{+2.4}_{-2.6}$ Gpc, corresponding to a redshift of $0.82^{+0.28}_{-0.34}$. The inferred rate of mergers similar to GW190521 is $0.13^{+0.30}_{-0.11}\ \mathrm{Gpc^{-3}\ yr^{-1}}$.

*Introduction.*—Advanced LIGO [1] and Advanced Virgo [2] have demonstrated a new means to observe the Universe through the detection of gravitational waves (GWs). In their first two observing runs (O1 and O2), the LIGO Scientific Collaboration and the Virgo Collaboration (LVC) have reported the detection of GWs from 10 binary black hole (BH) mergers, and a binary neutron star inspiral [3,4]. The third observing run (O3) started on April 1, 2019, and was suspended on March 27, 2020; numerous public alerts pertaining to possible detections have been sent to the astronomical community [5], with three confirmed detections [6–8].

The discovery of GW150914 [9] and subsequent events has revealed a population of binary BHs with total masses between ∼19 and 84 $M_\odot$, with component masses ranging from ∼8 to 50 $M_\odot$ [3]. Signals consistent with heavier BHs (e.g., 170817 + 03:02:46UTC) have also been reported in [10–12], albeit with a non-negligible chance of having an instrumental origin. For the parametrized population models considered in [13] it was inferred that no more than 1% of primary BH masses in merging binaries are greater than 45 $M_\odot$.

In this Letter we expand this mass range with the confident detection of GW190521, a GW signal consistent with a binary BH merger of total mass ∼150 $M_\odot$, leaving behind a ∼140 $M_\odot$ remnant. Waveform models for quasicircular binary BHs indicate that a precessing orbital plane is slightly favored over a fixed plane. The observation of the ringdown signal from the remnant BH provides estimates for the final mass and spin that are consistent with those from the full waveform analysis.

It is predicted that stars with a helium core mass in the range of ∼32–64 $M_\odot$ are subject to pulsational pair instability, leaving behind remnants with mass less than ∼65 $M_\odot$. Stars with helium core mass in the range ∼64–135 $M_\odot$ would be susceptible to pair instability and leave no compact remnant, while stars with helium mass $\gtrsim$135 $M_\odot$ are thought to directly collapse to intermediate mass BHs (IMBHs) [14–19]. The LVC O1-O2 observations are consistent with the prevention of heavy BH formation by pair-instability supernova (PISN) [13]. For GW190521, the mass of the heavier binary component has a high probability to be within the PISN mass gap [17,20–22]. In dense stellar systems or active galactic nuclei disks, BHs with mass in the PISN gap might form via hierarchical coalescence of smaller BHs [23–28], or via direct collapse of a stellar merger between an evolved star and a main sequence companion [29,30].

BHs of mass $10^2$–$10^5$ $M_\odot$, more massive than stellar mass BHs and lighter than supermassive BHs (SMBHs), are traditionally designated IMBHs [31–33]. A conclusive observation of these objects has thus far remained elusive, despite indirect evidence. These include observations of central BHs in galaxies, kinematical measurements of massive star clusters, scaling relations between the mass of the central SMBH and their host galaxies, and the mass range of globular clusters [34]. The LVC has also previously searched for binaries of IMBHs explicitly in their GW data, for example [35–37], obtaining null results and establishing an upper limit of 0.2 $\mathrm{Gpc^{-3}\,yr^{-1}}$ on their

*Full author list given at the end of the Letter.

coalescence rate [38]. The remnant of GW190521 fulfills the above definition of an IMBH.

GW190521 was detected by searches for quasicircular binary coalescences, and there is no evidence in the data for significant departures from such a signal model. However, for any transient with high inferred masses, there are few cycles observable in ground-based detectors, and therefore alternative signal models may also fit the data. This is further addressed in the companion paper [39] that also provides details about physical parameter estimation, and the astrophysical implications of the observation of GWs from this massive system.

*Observation.*—On May 21, 2019 at 03:02:29 UTC, the LIGO Hanford (LHO), LIGO Livingston (LLO), and Virgo observatories detected a coincident transient signal. A matched-filter search for compact binary mergers, PyCBC Live [40,41,42], reported the transient with a network signal-to-noise ratio (SNR) of 14.5 and a false-alarm rate of 1 in 8 yr, triggering the initial alert. A weakly modeled transient search based on coherent wave burst (CWB) [43] in its IMBH search configuration [35] reported a signal with a network SNR of 15.0 and a false-alarm rate lower than 1 in 28 yr. Two other matched-filter pipelines, SPIIR [44] and GstLAL [45], found consistent candidates albeit with higher false-alarm rates. The identification, localization, and classification of the transient as a binary BH merger were reported publicly within ≈6 min, with the candidate name S190521g [46,47].

A second significant GW trigger occurred on the same day at 07:43:59 UTC, S190521r [48]. Despite the short time separation, the inferred sky positions of GW190521 and S190521r are disjointed at high confidence, and so the events are not related by gravitational lensing. Further discussions pertaining to gravitational lensing and GW190521 are presented in the companion paper [39].

GW190521, shown in Fig. 1, is a short transient signal with a duration of approximately 0.1 s and around four cycles in the frequency band 30–80 Hz. A frequency of 60 Hz at the signal peak and the assumption that the source is a compact binary merger imply a massive system.

*Data.*—The LIGO and Virgo strain data are conditioned prior to their use in search pipelines and parameter estimation analyses. During online calibration of the data [53], narrow spectral features (lines) are subtracted using auxiliary witness sensors. Specifically, we remove from the data the 60 Hz U.S. mains power signature (LIGO), as well as calibration lines (LIGO and Virgo) that are intentionally injected into the detectors to measure the instruments' responses. During online calibration of Virgo data, broadband noise in the 40–1000 Hz frequency range is subtracted from the data [54]. The noise-subtracted data produced by the online calibration pipelines are used by online search pipelines and initial parameter estimation analyses.

Subsequent to the subtraction conducted within the online calibration pipeline, we perform a secondary offline subtraction [55] on the LIGO data with the goal of removing nonlinear sidebands around the U.S. mains power frequency, caused by low frequency modulation of the 60 Hz noise coupling. Since the subtraction of these sidebands is not expected to significantly improve the sensitivity of search algorithms, it is only used in offline parameter estimation of GW190521. Although GW190521 demonstrates a peak frequency of about 60 Hz, there is no evidence that the power mains contribute coherent power to the recovered signal. Voltage monitors and magnetometers installed at each LIGO site show no evidence of significant power fluctuations at the time of the event. These sensors are more sensitive to mains voltage transients than the interferometers are, detecting voltage fluctuations that are much smaller than those that produce transient noise in the strain data.

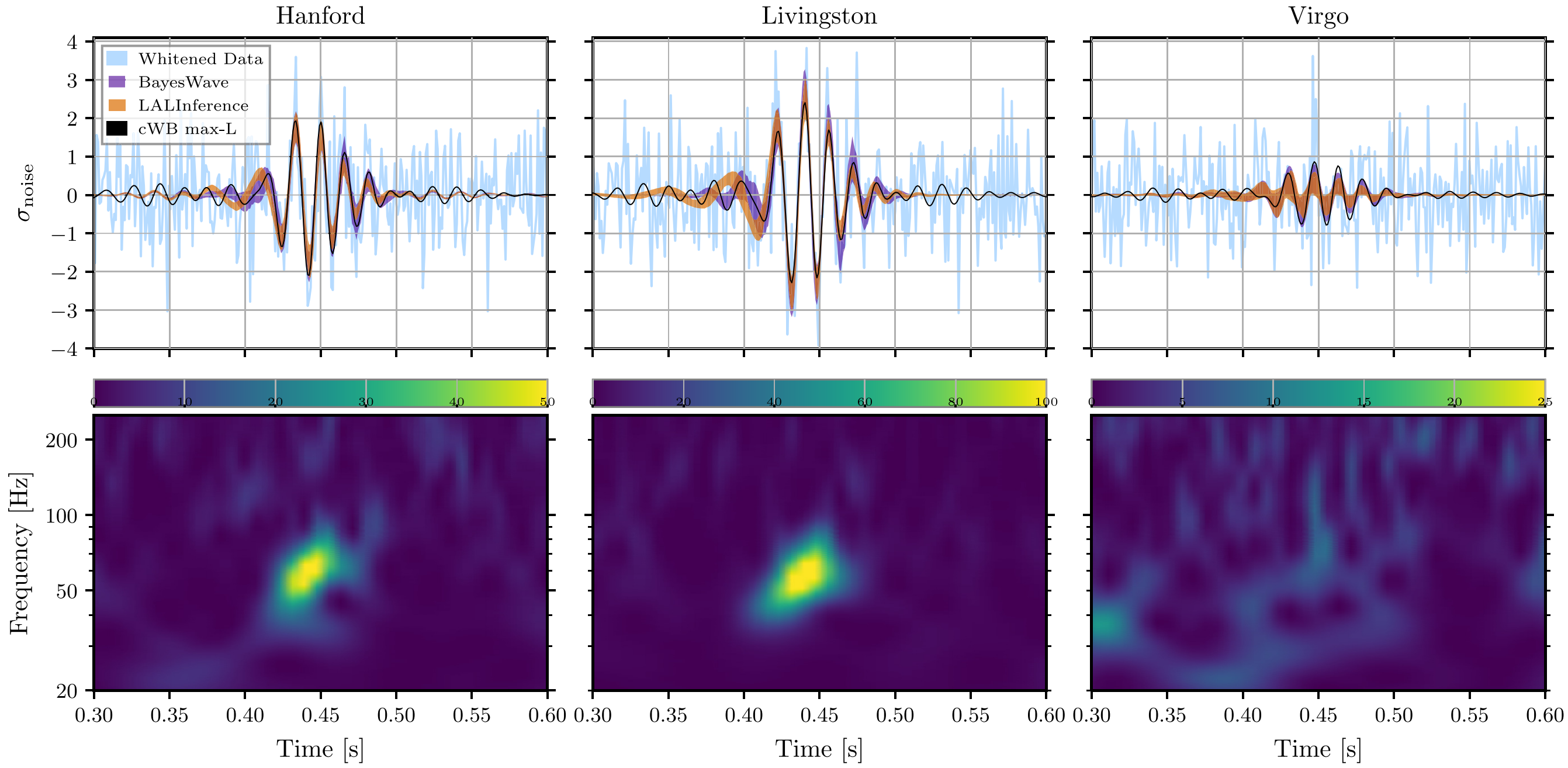

FIG. 1. The GW event GW190521 observed by the LIGO Hanford (left), LIGO Livingston (middle), and Virgo (right) detectors. Times are shown relative to May 21, 2019 at 03:02:29 UTC. The top row displays the time-domain detector data after whitening by each instrument's noise amplitude spectral density (light blue lines); the point estimate waveform from the CWB search [43] (black lines); the 90% credible intervals from the posterior probability density functions of the waveform time series, obtained via Bayesian inference (LALInference [49]) with the NRSur7dq4 binary BH waveform model [50] (orange bands), and with a generic wavelet model (BayesWave [51], purple bands). The ordinate axes are in units of noise standard deviations. The bottom row displays the time-frequency representation of the whitened data using the $Q$ transform [52].

At the time of GW190521, the LHO, LLO, and Virgo detectors were observing in their nominal operational O3 state. Low-latency data quality checks [56] did not indicate any transient noise in the vicinity of this event. Four minutes after GW190521, LHO microphones recorded the sound of a nearby helicopter, which also affected the GW strain data. This noise does not impact the confidence of the detection and the affected data are not used for parameter estimation. More thorough analyses performed at higher latency [3,56] find no evidence that GW190521 is due to, or influenced by, instrumental or environmental noise.

To further confirm that GW190521 is not a noise artifact, we followed the treatment in [3,56] and investigated potential sources of nonstationary noise typically found in the same frequency band measured for GW190521. The false-alarm rates calculated by the search pipelines estimate the rate of random coincidences of all glitches from the analysis period. Subsequent evaluation of the background noise relevant to an event does not change its calculated false-alarm rate, but serves solely as an event validation procedure. During local daytime hours, the LLO detector exhibits nonstationary noise that is consistent with scattered light due to excess ground motion in the 1–10 Hz band [3]. It produces a variation of the detector noise below 50 Hz, appearing as a periodic sequence of short duration transients. A similar type of noise is also observed in the LHO detector but at significantly lower rate. GW190521 was detected at 03:02:29 UTC, at which time the 1–10 Hz ground motion was low and the GW strain data are not exhibiting the characteristic nonstationarity associated with excessive scattered light. Both detectors also exhibit populations of short duration, band-limited transients (blip glitches) [57,58], which often demonstrate a characteristic frequency of ∼50 Hz. These transients are not found in coincidence between the LHO and LLO detectors (except by random occurrence) and GW190521 does not demonstrate the typical frequency-domain power distribution of blip glitches.

*Detection significance.*—After the identification by the low-latency analyses described above, GW190521 was also identified by offline analyses. These analyze strain data with improved calibration and updated data-quality vetoes, which are not available in low latency and hence update the low-latency results. The offline analyses use the CWB [43,59,60], GstLAL [45,61–65], and PyCBC [40,66–71] pipelines. CWB searches for short transient signals with minimal assumptions on their waveform. GstLAL and PyCBC search for coalescences of compact objects using matched filtering with banks of quasicircular, quadrupolar-mode-only, nonprecessing templates [72–78].

We performed the offline CWB analyses (see the Supplemental Material [79]) using two detector configurations: one restricted to the LIGO detectors, and one including Virgo as well. These two analyses identified GW190521 with network SNRs of 14.4 and 14.7, respectively, and with event parameters well within the limits defined by the analysis selection cuts. The LIGO-only analysis was used to establish the false-alarm rate for GW190521. The analysis including Virgo produced the waveform reconstruction. The GW190521 false-alarm rate was estimated from the analysis of time-shifted LIGO data. The background is equivalent to 9800 yr of observation and contains only two events ranked higher than GW190521, both consistent with random coincidences of short duration (∼1 cycle) glitches observed in the LIGO frequency band 20–100 Hz. The estimated background results in a false-alarm rate of 1 in 4900 yr for GW190521, which constitutes a confident detection of a GW transient.

The offline analysis conducted by GstLAL (see the Supplemental Material [79]) identified GW190521 with a network matched-filter SNR of 14.7 and a false-alarm rate of 1 in 829 yr. The large difference in GstLAL significance reported by its online and offline configurations is due to an improvement in the template bank during O3 that greatly enhanced GstLAL's sensitivity to mergers of high-mass compact objects.

The offline analysis performed by PyCBC (see Supplemental Material [79]) identified GW190521 with a network matched-filter SNR of 12.6 and a false-alarm rate of 1 in 0.94 yr. The smaller SNR and relatively high false-alarm rate are due to the sparseness of PyCBC's template bank in the parameter region of GW190521, coupled with the fact that instrumental transients cause different high-mass templates to produce very different rates of high-SNR triggers. The Supplemental Material [79] describes PyCBC's response to GW190521 in greater detail.

The most massive binary BH merger previously reported by the LVC, GW170729, had the same ordering of significances in CWB, GstLAL, and PyCBC as GW190521, and a simulation campaign showed that larger significances in CWB for such heavy BH mergers are not uncommon [3]. Matched-filter searches based on quasicircular nonprecessing templates and CWB have also been compared using broader simulations of heavy BH mergers, including precession and higher-order multipole moments, also concluding that CWB is often more sensitive [80,81]. We performed a similar simulation campaign for GW190521 in order to further understand the different significances. We simulated thousands of signals compatible with the parameters inferred for the event under the assumption of a quasicircular BH merger, using the NRSur7dq4 waveform model described in the next section, which includes precession and higher-order multipole moments. The simulated sources have merger times distributed uniformly over several days surrounding GW190521, so as to sample many different realizations of the detector noise. The right ascensions have been correspondingly corrected in order to cancel the effect of Earth's rotation, which would lead to different projections of the strain polarizations on the detectors. We added the signals into the data surrounding the event, reran the search pipelines with the same configuration used for the offline analysis, and counted the number of signals recovered by each pipeline. CWB, GstLAL, and PyCBC recovered, respectively, 36%, 45%, and 11% of the simulated signals at a false-alarm rate better than 1 in 4900 yr. The fraction of signals found at a false-alarm rate in CWB better than 1 in 4900 yr and a false-alarm rate in PyCBC worse than 1 in 0.94 yr is 2.7%, which is small but not negligible. The fraction found at a false-alarm rate in CWB better than 1 in 4900 yr and a false-alarm rate in GstLAL worse than 1 in 829 yr is 7.8%.

We conclude that the outputs of CWB, GstLAL, and PyCBC are fully consistent with expectations for a quasicircular binary merger signal with the parameters of GW190521. The reported false-alarm rates do not include a trials factor for the number of analyses performed. If one were to choose a single representative false-alarm rate, one should use the CWB rate multiplied by a trials factor of 3, resulting from the conservative assumption [37,38] that CWB, GstLAL, and PyCBC are equally sensitive and statistically independent. The resulting rate would still point to a significant detection.

*Astrophysical source.*—GW190521 is qualitatively different from previous detections [3,6–8] due to the small number of cycles and maximum frequency in the sensitive band of the detectors. Hence, its astrophysical interpretation as a quasicircular compact binary merger warrants more discussion than previous events. Alternative scenarios, such as an eccentric collision [82], become more relevant and are discussed in the companion paper [39]. Nevertheless, the quasicircular BH merger scenario remains the most plausible and we will proceed under this assumption in the rest of this Letter.

We performed Bayesian parameter inference on GW190521 using three waveform models for quasicircular binary BHs including the effects of higher order multipole moments and precession. These are the numerical relativity surrogate model NRSur7dq4 [50], the effective-one-body model SEOBNRv4PHM [83,84], and the phenomenological model IMRPhenomPv3HM [85]. To compute the evidence for the presence of higher-order modes, orbital precession and nonzero spin, we also compared the data with the aforementioned models after removing these effects from the models. We analyzed 8 s of data around the time of GW190521. We impose uniform priors on the redshifted component masses, on the individual spin magnitudes and on the square of the luminosity distance. We have checked that imposing an uniform-in-co-moving-volume prior changes the results by less than 1%. We impose an isotropic prior on the source and the spin orientations. We produce posterior distributions marginalized over calibration uncertainties. For the NRSur7dq4 and IMRPhenomPHM runs, we made use of the LALInference software package [49] while the SEOBNRv4PHM runs were done using the RIFT algorithm [86]. We find that despite differences in how these waveform models are computed, and the fact that we needed to sample over parameters outside their calibration regions [87], all yield broadly consistent results [39]. In addition, direct comparison of the data to numerical relativity simulations [88–90], using the RIFT algorithm, yields consistent results. In the following we quote results obtained using the NRSur7dq4 model. This choice is motivated by this being the only model that has been calibrated to numerical simulations of precessing BH binaries. The NRSur7dq4 results are summarized in Table I. Results for all three models are presented in the companion paper [39].

TABLE I. Parameters of GW190521 according to the NRSur7dq4 waveform model. We quote median values with 90% credible intervals that include statistical errors.

| Parameter | |
|---|---|
| Primary mass | $85^{+21}_{-14}\ M_\odot$ |
| Secondary mass | $66^{+17}_{-18}\ M_\odot$ |
| Primary spin magnitude | $0.69^{+0.27}_{-0.62}$ |
| Secondary spin magnitude | $0.73^{+0.24}_{-0.64}$ |
| Total mass | $150^{+29}_{-17}\ M_\odot$ |
| Mass ratio ($m_2/m_1 \leq 1$) | $0.79^{+0.19}_{-0.29}$ |
| Effective inspiral spin parameter ($\chi_{\rm eff}$) | $0.08^{+0.27}_{-0.36}$ |
| Effective precession spin parameter ($\chi_{\rm p}$) | $0.68^{+0.25}_{-0.37}$ |
| Luminosity Distance | $5.3^{+2.4}_{-2.6}$ Gpc |
| Redshift | $0.82^{+0.28}_{-0.34}$ |
| Final mass | $142^{+28}_{-16}\ M_\odot$ |
| Final spin | $0.72^{+0.09}_{-0.12}$ |
| $P\ (m_1 < 65\ M_\odot)$ | 0.32% |
| $\log_{10}$ Bayes factor for orbital precession | $1.06^{+0.06}_{-0.06}$ |
| $\log_{10}$ Bayes factor for nonzero spins | $0.92^{+0.06}_{-0.06}$ |
| $\log_{10}$ Bayes factor for higher harmonics | $-0.38^{+0.06}_{-0.06}$ |

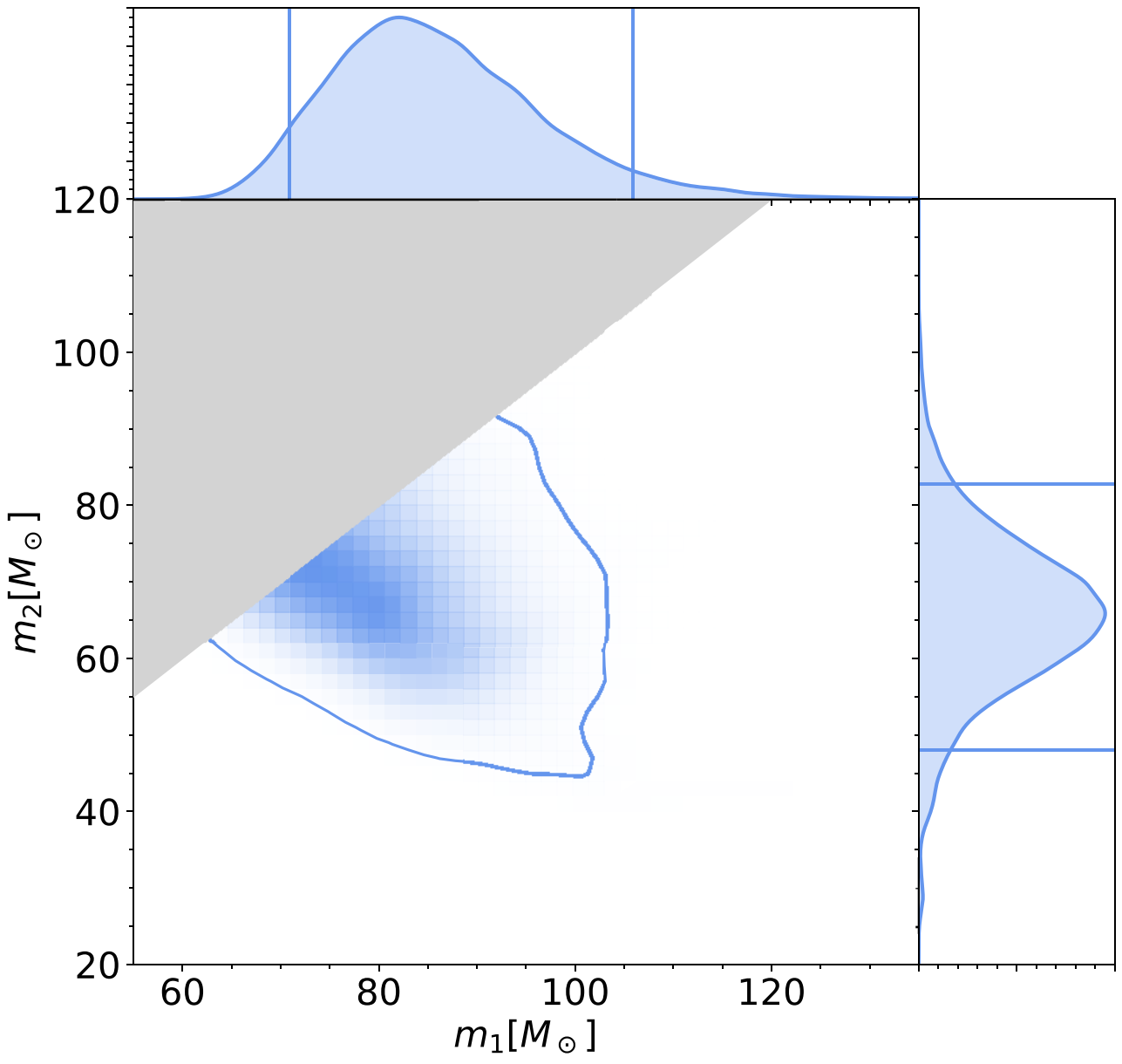

FIG. 2. Posterior distributions for the progenitor masses of GW190521 according to the NRSur7dq4 waveform model. The 90% credible regions are indicated by the solid contour in the joint distribution and by solid vertical and horizontal lines in the marginalized distributions.

Figure 2 shows our estimated 90% credible regions for the individual masses of GW190521. We estimate individual components with $(m_1, m_2) = (85^{+21}_{-14}, 66^{+17}_{-18})\ M_\odot$ and a total mass $150^{+29}_{-17}\ M_\odot$. This makes GW190521 the most massive binary BH observed to date, as expected from its short duration and low peak frequency. To quantify compatibility with the PISN mass gap, we find the probability of the primary component being below 65 $M_\odot$ to be 0.32%. The estimated mass and dimensionless spin magnitude of the remnant object are $M_\mathrm{f} = 142^{+28}_{-16}\ M_\odot$ and $\chi_\mathrm{f} = 0.72^{+0.09}_{-0.12}$ respectively. The posterior for $M_\mathrm{f}$ shows no support below 100 $M_\odot$, making the remnant the first conclusive direct observation of an IMBH.

The left panel of Fig. 3 shows the posterior distributions for the magnitude and tilt angle of the individual spins, measured at a reference frequency of 11 Hz. All pixels in this plot have equal prior probability. While we obtain posteriors with strong support at the $\chi = 1$ limit imposed by cosmic censorship [91], these also show non-negligible support for zero spin magnitudes. In addition, the maximum posterior probability corresponds to large angles between the spins and the orbital angular momentum. Large spin magnitudes and tilt angles would lead to a strong spin-orbit coupling, causing the orbital plane to precess [92,93]. The impact of precession in a GW signal is commonly parametrized by the effective precession spin parameter $\chi_p$ [94,95] while the effective inspiral spin parameter $\chi_{\rm eff}$ parametrizes the impact of the spin components aligned with the orbital angular momentum [96–99]. The right panel of Fig. 3 shows the corresponding posterior distributions. We estimate $\chi_{\rm eff} = 0.08^{+0.27}_{-0.36}$ and $\chi_p = 0.68^{+0.25}_{-0.37}$.

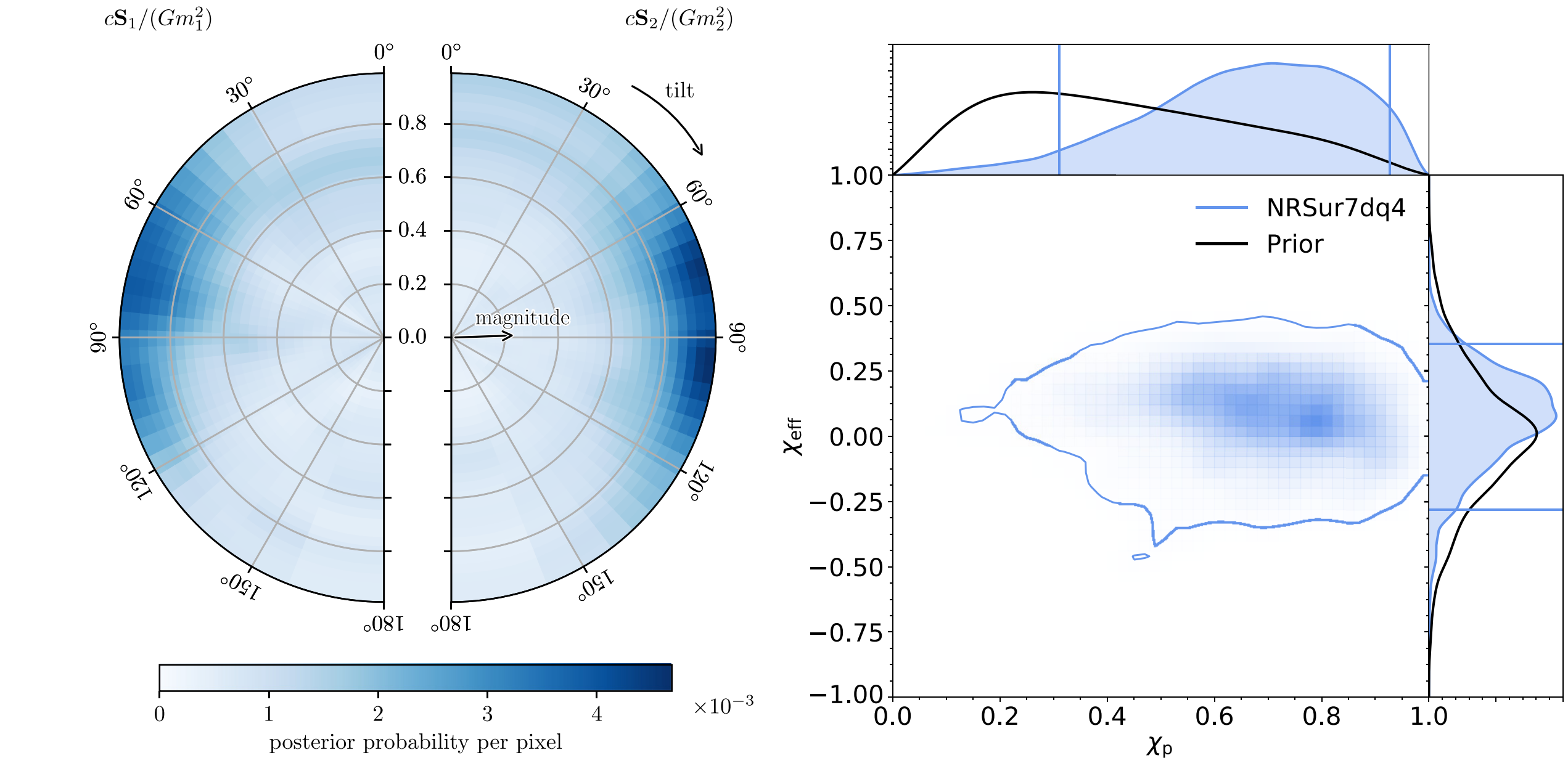

FIG. 3. Left: posterior distribution for the individual spins of GW190521 according to the NRSur7dq4 waveform model. The radial coordinate in the plot denotes the dimensionless spin magnitude, while the angle denotes the spin tilt, defined as the angle between the spin and the orbital angular momentum of the binary at reference frequency of 11 Hz. A tilt of 0° indicates that the spin is aligned with the orbital angular momentum. A nonzero magnitude and a tilt away from 0° and 180° imply a precessing orbital plane. All bins have equal prior probability. Right: posterior distributions for the effective spin and effective in-plane spin parameters. The 90% credible regions are indicated by the solid contour in the joint distribution, and by solid vertical and horizontal lines in the marginalized distributions. The large density for tilts close to 90° leads to large values for $\chi_\mathrm{p}$ and low values for $\chi_\mathrm{eff}$.

We evaluated the Bayesian evidence for both a precessing orbital plane and nonzero spin magnitudes by performing model selection with models omitting precession and spins. We obtain a $\log_{10}$ Bayes factor of $1.06^{+0.06}_{-0.06}$ in favor of precessing versus nonprecessing spins and $0.92^{+0.06}_{-0.06}$ in favor of nonzero spin magnitudes versus zero magnitudes. This indicates a weak preference for both a spinning BBH and a precessing orbital plane, consistent with the large uncertainty in the spin parameters and the fact that the final spin, $\chi_f = 0.72^{+0.09}_{-0.12}$, is consistent with a nonspinning BBH [3]. Future analyses of GW190521 with improved waveform models and more informed population priors may well shift the maximum probability to other regions of the spin parameter space.

We estimate the luminosity distance of GW190521 to be $5.3^{+2.4}_{-2.6}$ Gpc, corresponding to a redshift of $0.82^{+0.28}_{-0.34}$, assuming a $\Lambda$CDM cosmology with Hubble parameter $H_0 = 67.9\ {\rm km\,s^{-1}\ Mpc^{-1}}$ [100]. Figure 4 shows the joint posterior distribution for the luminosity distance and the inclination angle between the total angular momentum of the binary and the line of sight, $\theta_{JN}$. We constrain $\sin(\theta_{JN}) < 0.79$ at the 90% credible level. Signals emitted at such inclinations are dominated by the quadrupolar $(2, \pm 2)$ modes [101–104]. Indeed, we obtain a $\log_{10}$ Bayes factor of $-0.38^{+0.06}_{-0.06}$ disfavoring the presence of higher order multipole moments in the data. Despite this fact, as described in [105], models that include higher modes still lead to more precise estimates of the distance and inclination of the source. The reason is that higher modes are more prominent in signals with large inclination angles, especially when the signal is dominated by the merger and ringdown portions, thereby allowing us to discard those angles [103,104,106,107].

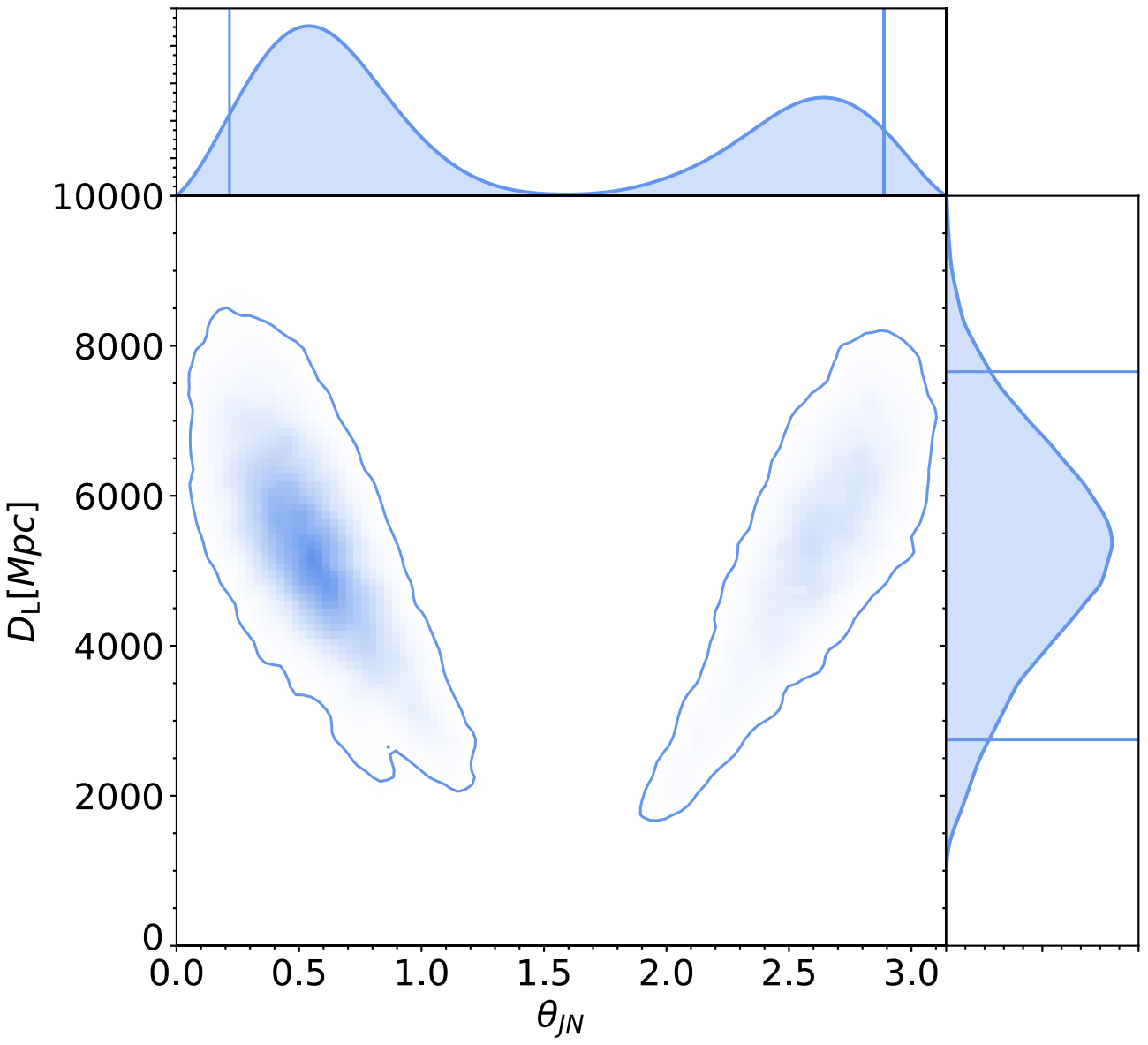

FIG. 4. Posterior distributions for the luminosity distance and the inclination angle of GW190521, according to the NRSur7dq4 waveform model. The inclination angle indicates the angle between the line-of-sight and the total angular momentum of the binary. For nonprecessing binaries, this is equal to the angle between the orbital angular momentum and the line of sight. We find the total angular momentum is likely to be closer to the line of sight than to the orthogonal direction. The solid lines and the central contour denote 90% credible regions.

Given that GW190521 has only a few observable cycles, a barely observable inspiral, and shows no evidence for higher-order modes, we investigate what aspects of the signal can lead to a slight evidence for nonzero spins and a precessing orbital plane. To do this, we compare the posterior sample waveforms obtained by the analyses including and omitting precession. We find that the most prominent effect of precession is a slight amplitude suppression of the lowest-frequency part of the waveform, consistent with the amplitude modulation typically associated with precession [82,108,109]. Meanwhile, the spin degrees of freedom that most affect $\chi_f$ are encapsulated in $\chi_{\rm eff}$ and not $\chi_p$ [110,111], so it is unlikely that our measurement of the final mass and spin informs $\chi_p$. Hence, the shift of the posteriors towards large spin magnitudes and tilt angles is more likely caused by the dynamics immediately prior to the merger, rather than postmerger features in the data.

Similarly, we have investigated how information about the mass ratio is being retrieved. We partially attribute this to the measurement of the remnant spin $\chi_f$ from information in the ringdown phase. This constrains the possible values of $q$ that can give rise to the measured $\chi_f$. In addition, the frequency at the signal peak amplitude [112,113], and the phase and amplitude evolution of the (suppressed) premerger signal can further constrain the measurement.

With only one such system so far confirmed, uncertainties on the formation channel and corresponding merger rate are necessarily very large. Under the simplifying assumption that the component masses and spins of GW190521 are representative of a population of merging binaries, we estimate a merger rate $0.13^{+0.30}_{-0.11}\ {\rm Gpc^{-3}\,yr^{-1}}$ [39], consistent with the prior upper bounds set in [38].

*Waveform reconstruction.*—GW190521 waveform reconstructions are obtained through a templated LALInference analysis [49], and two signal-agnostic analyses, CWB [43,114], and BayesWave [51,115]. Both signal-agnostic analyses reconstruct signal waveforms as a linear combination of wavelets: CWB obtains point estimate waveforms with the constrained maximum likelihood method while BayesWave reconstructs waveforms by drawing posterior samples from an unmodeled Bayesian analysis. Figure 1 shows broad agreement between the waveform reconstructions.

For a quantitative comparison of the CWB point estimate waveform $w$ and the template $h$, we calculate the overlap, or match $(w|h)/\sqrt{(w|w)(h|h)}$, where $(w|h)$ denotes the noise-weighted network inner product [116]. We randomly draw signals from the templated inference analysis, inject these into data surrounding GW190521, and reconstruct the injections with CWB. The overlaps between the simulated signals and the corresponding CWB reconstructions define the null distribution, which takes into account the waveform reconstruction errors and fluctuations of the detector noise. The median and 90% confidence interval for the null distribution are $0.93^{+0.03}_{-0.06}$. The overlap between the CWB point estimate for GW190521 and the maximum-likelihood NRSur7dq4 template is 0.89 and is consistent with the null distribution.

The overlap [115] between the median BayesWave waveform and the maximum likelihood NRSur7dq4 template is 0.93. A signal residual test is performed by subtracting the maximum likelihood NRSur7dq4 template from the data and then searching for a residual signal using BayesWave [117]. The residual search result is compared to the distribution found from the analysis of the off-source data surrounding the event. This comparison results in a $p$ value (as first described in [118]) of 0.4, indicating that the residual is fully consistent with noise.

*Black hole ringdown.*—We analyzed the ringdown portion of GW190521 using a damped sinusoid to fit the least-damped ringdown mode [119,120]. Starting 12.7 ms after the peak of the complex strain [corresponding to $\sim t_{\rm peak} + 10G(1+z)M_{\rm f}/c^3$ in units of the redshifted remnant mass $(1+z)M_{\rm f}$ [121], using median values from the NRSur7dq4 approximant], the analysis estimates a frequency $f = 66^{+4}_{-3}$ Hz and damping time $\tau = 19^{+9}_{-7}$ ms, with a Bayes factor between signal and noise of $\log_{10}(B_{s/n}) = 25.45 \pm 0.02$. By imposing predictions of perturbation theory on the frequency of the GW emission [122] we infer the final redshifted mass and dimensionless spin to be $(1+z)M_{\rm f} = 252^{+63}_{-64}\ M_\odot$ and $\chi_f = 0.65^{+0.22}_{-0.48}$. All quoted values correspond to median and 90% credible intervals. The grey contour in Fig. 5 shows the corresponding posterior two-dimensional 90% credible region. Accounting for redshift, these results are consistent with the full-waveform analysis when using NR fits to predict the remnant quantities [50,110,122–125]. The corresponding 90% credible region is shown in blue in the same Fig. 5. Additional detailed investigations are reported in the companion paper [39].

*Summary.*—GW190521 is a short duration signal consistent with a binary BH merger. According to state of the art models for quasicircular binaries, the progenitor BHs show mild evidence for nonzero spins and a precessing orbit, and the heavier component mass $85^{+21}_{-14}\ M_\odot$ sits in the PISN mass gap. The merger left behind a remnant with a final mass of $142^{+28}_{-16}\ M_\odot$, making this a direct observation of the formation of an IMBH. The large individual and total masses of GW190521, and the low likelihood that the primary originated from a stellar collapse given theoretical constraints on supernova physics, strongly suggest a different formation channel from BH binaries previously reported. The remnant ringdown signal is compatible with the full waveform analysis and general relativity. The short duration of GW190521 also invites other interpretations for the source. Further details on the properties of GW190521 are discussed in the companion paper, together with its astrophysical implications and possible formation channels [39]. As the low frequency sensitivity improves for Advanced LIGO and Advanced Virgo [126] further massive binary BH events should be observed. Third-generation ground-based GW detectors [127–129] and LISA [130] will be important instruments to study these systems. An event like GW190521 may be observable by both LISA and ground-based detectors, and observing the earlier inspiral is essential to improve our understanding of the source [131–134].

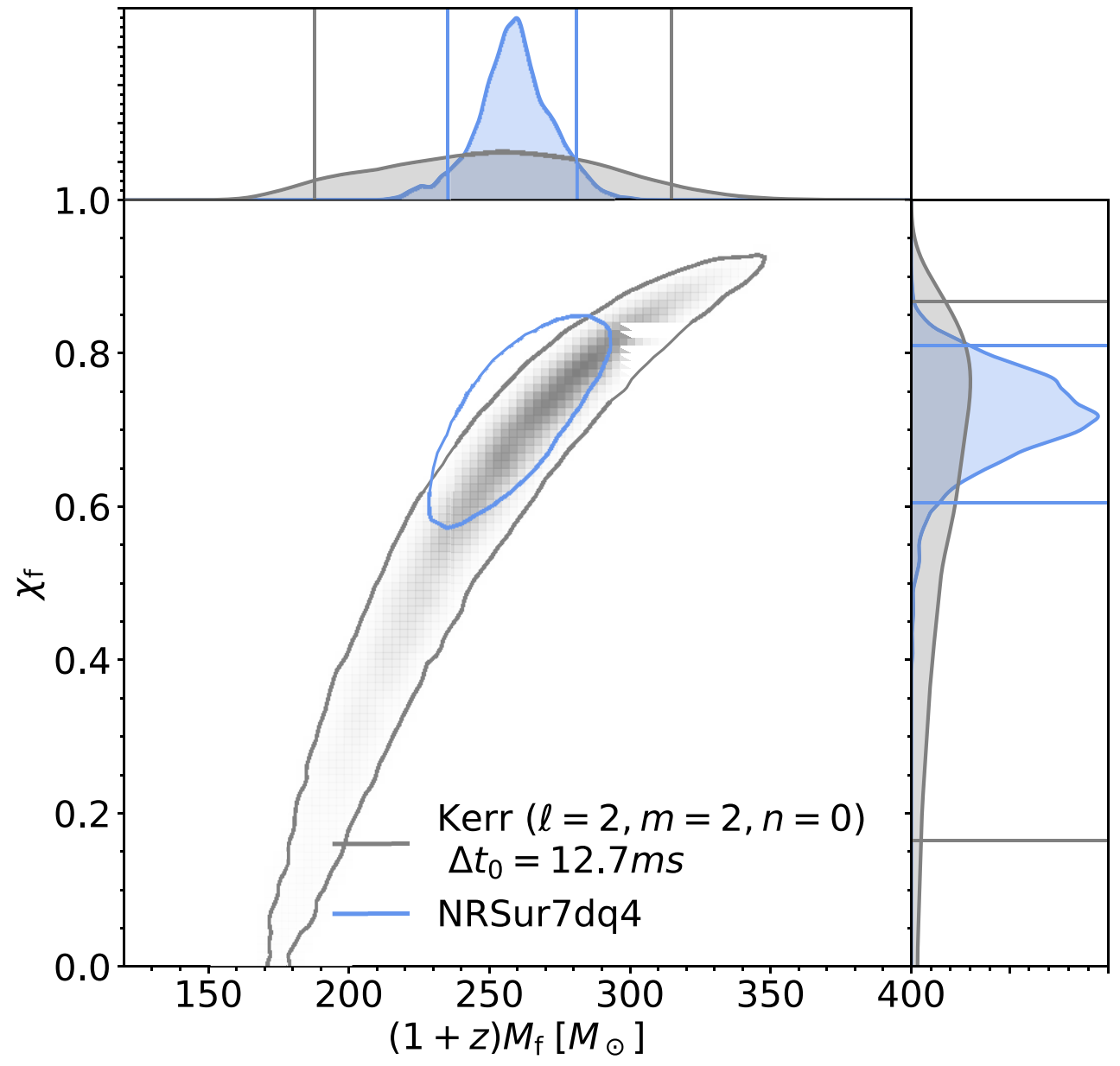

FIG. 5. Redshifted remnant mass and spin inferred from the least-damped $\ell = m = 2$ ringdown mode. The analysis was carried out 12.7 ms [$\sim 10G(1+z)M_{\rm f}/c^3$] after the reference time $t^{\rm H}_{\rm peak} = 1\,242\,442\,967.4306$ for the Hanford detector (appropriately time shifted in the other detectors assuming the maximum likelihood value on the sky position inferred from the NRSur7dq4 approximant). The blue contour represents the 90% credible region of the prediction from the full-waveform analysis.

Strain data from the LIGO and Virgo detectors associated with GW190521, and supporting data for this Letter, can be found at [135].

The authors gratefully acknowledge the support of the United States National Science Foundation (NSF) for the construction and operation of the LIGO Laboratory and Advanced LIGO as well as the Science and Technology Facilities Council (STFC) of the United Kingdom, the Max-Planck-Society (MPS), and the State of Niedersachsen/Germany for support of the construction of Advanced LIGO and construction and operation of the GEO600 detector. Additional support for Advanced LIGO was provided by the Australian Research Council. The authors gratefully acknowledge the Italian Istituto Nazionale di Fisica Nucleare (INFN), the French Centre National de la Recherche Scientifique (CNRS) and the Netherlands Organization for Scientific Research, for the construction and operation of the Virgo detector and the creation and support of the EGO consortium. The authors also gratefully acknowledge research support from these agencies as well as by the Council of Scientific and Industrial Research of India, the Department of Science and Technology, India, the Science & Engineering Research Board (SERB), India, the Ministry of Human Resource Development, India, the Spanish Agencia Estatal de Investigación, the Vicepresidència i Conselleria d'Innovació, Recerca i Turisme and the Conselleria d'Educació i Universitat del Govern de les Illes Balears, the Conselleria d'Innovació, Universitats, Ciència i Societat Digital de la Generalitat Valenciana and the CERCA Programme Generalitat de Catalunya, Spain, the National Science Centre of Poland, the Swiss National Science Foundation (SNSF), the Russian Foundation for Basic Research, the Russian Science Foundation, the European Commission, the European Regional Development Funds (ERDF), the Royal Society, the Scottish Funding Council, the Scottish Universities Physics Alliance, the Hungarian Scientific Research Fund (OTKA), the French Lyon Institute of Origins (LIO), the Belgian Fonds de la Recherche Scientifique (FRS-FNRS), Actions de Recherche Concertées (ARC) and Fonds Wetenschappelijk Onderzoek—Vlaanderen (FWO), Belgium, the Paris Île-de-France Region, the National Research, Development and Innovation Office Hungary (NKFIH), the National Research Foundation of Korea, Industry Canada and the Province of Ontario through the Ministry of Economic Development and Innovation, the Natural Science and Engineering Research Council Canada, the Canadian Institute for Advanced Research, the Brazilian Ministry of Science, Technology, Innovations, and Communications, the International Center for Theoretical Physics South American Institute for Fundamental Research (ICTP-SAIFR), the Research Grants Council of Hong Kong, the National Natural Science Foundation of China (NSFC), the Leverhulme Trust, the Research Corporation, the Ministry of Science and Technology (MOST), Taiwan and the Kavli Foundation. The authors gratefully acknowledge the support of the NSF, STFC, INFN, and CNRS for provision of computational resources. We thank the referees for useful comments that have improved the manuscript. In addition to the software cited earlier, PESUMMARY [136] was used to produce the publicly released samples and MATPLOTLIB [137] was used for plotting.

*Note added.*—Recently, a candidate optical counterpart to GW190521 was reported [138].

R. Abbott,[1] T. D. Abbott,[2] S. Abraham,[3] F. Acernese,[4,5] K. Ackley,[6] C. Adams,[7] R. X. Adhikari,[1] V. B. Adya,[8] C. Affeldt,[9,10] M. Agathos,[11,12] K. Agatsuma,[13] N. Aggarwal,[14] O. D. Aguiar,[15] A. Aich,[16] L. Aiello,[17,18] A. Ain,[3] P. Ajith,[19] S. Akcay,[11] G. Allen,[20] A. Allocca,[21] P. A. Altin,[8] A. Amato,[22] S. Anand,[1] A. Ananyeva,[1] S. B. Anderson,[1] W. G. Anderson,[23]

S. V. Angelova,[24] S. Ansoldi,[25,26] S. Antier,[27] S. Appert,[1] K. Arai,[1] M. C. Araya,[1] J. S. Areeda,[28] M. Arène,[27] N. Arnaud,[29,30] S. M. Aronson,[31] K. G. Arun,[32] Y. Asali,[33] S. Ascenzi,[17,34] G. Ashton,[6] S. M. Aston,[7] P. Astone,[35] F. Aubin,[36] P. Aufmuth,[10] K. AultONeal,[37] C. Austin,[2] V. Avendano,[38] S. Babak,[27] P. Bacon,[27] F. Badaracco,[17,18] M. K. M. Bader,[39] S. Bae,[40] A. M. Baer,[41] J. Baird,[27] F. Baldaccini,[42,43] G. Ballardin,[30] S. W. Ballmer,[44] A. Bals,[37] A. Balsamo,[41] G. Baltus,[45] S. Banagiri,[46] D. Bankar,[3] R. S. Bankar,[3] J. C. Barayoga,[1] C. Barbieri,[47,48] B. C. Barish,[1] D. Barker,[49] K. Barkett,[50] P. Barneo,[51] F. Barone,[52,5] B. Barr,[53] L. Barsotti,[54] M. Barsuglia,[27] D. Barta,[55] J. Bartlett,[49] I. Bartos,[31] R. Bassiri,[56] A. Basti,[57,21] M. Bawaj,[58,43] J. C. Bayley,[53] M. Bazzan,[59,60] B. Bécsy,[61] M. Bejger,[62] I. Belahcene,[29] A. S. Bell,[53] D. Beniwal,[63] M. G. Benjamin,[37] J. D. Bentley,[13] F. Bergamin,[9] B. K. Berger,[56] G. Bergmann,[9,10] S. Bernuzzi,[11] C. P. L. Berry,[14] D. Bersanetti,[64] A. Bertolini,[39] J. Betzwieser,[7] R. Bhandare,[65] A. V. Bhandari,[3] J. Bidler,[28] E. Biggs,[23] I. A. Bilenko,[66] G. Billingsley,[1] R. Birney,[67] O. Birnholtz,[68,69] S. Biscans,[1,54] M. Bischi,[70,71] S. Biscoveanu,[54] A. Bisht,[10] G. Bissenbayeva,[16] M. Bitossi,[30,21] M. A. Bizouard,[72] J. K. Blackburn,[1] J. Blackman,[50] C. D. Blair,[7] D. G. Blair,[73] R. M. Blair,[49] F. Bobba,[74,75] N. Bode,[9,10] M. Boer,[72] Y. Boetzel,[76] G. Bogaert,[72] F. Bondu,[77] E. Bonilla,[56] R. Bonnand,[36] P. Booker,[9,10] B. A. Boom,[39] R. Bork,[1] V. Boschi,[21] S. Bose,[3] V. Bossilkov,[73] J. Bosveld,[73] Y. Bouffanais,[59,60] A. Bozzi,[30] C. Bradaschia,[21] P. R. Brady,[23] A. Bramley,[7] M. Branchesi,[17,18] J. E. Brau,[78] M. Breschi,[11] T. Briant,[79] J. H. Briggs,[53] F. Brighenti,[70,71] A. Brillet,[72] M. Brinkmann,[9,10] P. Brockill,[23] A. F. Brooks,[1] J. Brooks,[30] D. D. Brown,[63] S. Brunett,[1] G. Bruno,[80] R. Bruntz,[41] A. Buikema,[54] T. Bulik,[81] H. J. Bulten,[82,39] A. Buonanno,[83,84] R. Buscicchio,[13] D. Buskulic,[36] R. L. Byer,[56] M. Cabero,[9,10] L. Cadonati,[85] G. Cagnoli,[86] C. Cahillane,[1] J. Calderón Bustillo,[6] J. D. Callaghan,[53] T. A. Callister,[1] E. Calloni,[87,5] J. B. Camp,[88] M. Canepa,[89,64] K. C. Cannon,[90] H. Cao,[63] J. Cao,[91] G. Carapella,[74,75] F. Carbognani,[30] S. Caride,[92] M. F. Carney,[14] G. Carullo,[57,21] J. Casanueva Diaz,[21] C. Casentini,[93,34] J. Castañeda,[51] S. Caudill,[39] M. Cavaglià,[94] F. Cavalier,[29] R. Cavalieri,[30] G. Cella,[21] P. Cerdá-Durán,[95] E. Cesarini,[96,34] O. Chaibi,[72] K. Chakravarti,[3] C. Chan,[90] M. Chan,[53] K. Chandra,[97] S. Chao,[98] P. Charlton,[99] E. A. Chase,[14] E. Chassande-Mottin,[27] D. Chatterjee,[23] M. Chaturvedi,[65] K. Chatziioannou,[100,101] H. Y. Chen,[102] X. Chen,[73] Y. Chen,[50] H.-P. Cheng,[31] C. K. Cheong,[103] H. Y. Chia,[31] F. Chiadini,[104,75] R. Chierici,[105] A. Chincarini,[64] A. Chiummo,[30] G. Cho,[106] H. S. Cho,[107] M. Cho,[84] N. Christensen,[72] Q. Chu,[73] S. Chua,[79] K. W. Chung,[103] S. Chung,[73] G. Ciani,[59,60] P. Ciecielag,[62] M. Cieślar,[62] A. A. Ciobanu,[63] R. Ciolfi,[108,60] F. Cipriano,[72] A. Cirone,[89,64] F. Clara,[49] J. A. Clark,[85] P. Clearwater,[109] S. Clesse,[80] F. Cleva,[72] E. Coccia,[17,18] P.-F. Cohadon,[79] D. Cohen,[29] M. Colleoni,[110] C. G. Collette,[111] C. Collins,[13] M. Colpi,[47,48] M. Constancio Jr.,[15] L. Conti,[60] S. J. Cooper,[13] P. Corban,[7] T. R. Corbitt,[2] I. Cordero-Carrión,[112] S. Corezzi,[42,43] K. R. Corley,[33] N. Cornish,[61] D. Corre,[29] A. Corsi,[92] S. Cortese,[30] C. A. Costa,[15] R. Cotesta,[83] M. W. Coughlin,[1] S. B. Coughlin,[113,14] J.-P. Coulon,[72] S. T. Countryman,[33] P. Couvares,[1] P. B. Covas,[110] D. M. Coward,[73] M. J. Cowart,[7] D. C. Coyne,[1] R. Coyne,[114] J. D. E. Creighton,[23] T. D. Creighton,[16] J. Cripe,[2] M. Croquette,[79] S. G. Crowder,[115] J.-R. Cudell,[45] T. J. Cullen,[2] A. Cumming,[53] R. Cummings,[53] L. Cunningham,[53] E. Cuoco,[30] M. Curylo,[81] T. Dal Canton,[83] G. Dálya,[116] A. Dana,[56] L. M. Daneshgaran-Bajastani,[117] B. D'Angelo,[89,64] S. L. Danilishin,[9,10] S. D'Antonio,[34] K. Danzmann,[10,9] C. Darsow-Fromm,[118] A. Dasgupta,[119] L. E. H. Datrier,[53] V. Dattilo,[30] I. Dave,[65] M. Davier,[29] G. S. Davies,[120] D. Davis,[44] E. J. Daw,[121] D. DeBra,[56] M. Deenadayalan,[3] J. Degallaix,[22] M. De Laurentis,[87,5] S. Deléglise,[79] M. Delfavero,[68] N. De Lillo,[53] W. Del Pozzo,[57,21] L. M. DeMarchi,[14] V. D'Emilio,[113] N. Demos,[54] T. Dent,[120] R. De Pietri,[122,123] R. De Rosa,[87,5] C. De Rossi,[30] R. DeSalvo,[124] O. de Varona,[9,10] S. Dhurandhar,[3] M. C. Díaz,[16] M. Diaz-Ortiz Jr.,[31] T. Dietrich,[39] L. Di Fiore,[5] C. Di Fronzo,[13] C. Di Giorgio,[74,75] F. Di Giovanni,[95] M. Di Giovanni,[125,126] T. Di Girolamo,[87,5] A. Di Lieto,[57,21] B. Ding,[111] S. Di Pace,[127,35] I. Di Palma,[127,35] F. Di Renzo,[57,21] A. K. Divakarla,[31] A. Dmitriev,[13] Z. Doctor,[102] F. Donovan,[54] K. L. Dooley,[113] S. Doravari,[3] I. Dorrington,[113] T. P. Downes,[23] M. Drago,[17,18] J. C. Driggers,[49] Z. Du,[91] J.-G. Ducoin,[29] P. Dupej,[53] O. Durante,[74,75] D. D'Urso,[128,129] S. E. Dwyer,[49] P. J. Easter,[6] G. Eddolls,[53] B. Edelman,[78] T. B. Edo,[121] O. Edy,[130] A. Effler,[7] P. Ehrens,[1] J. Eichholz,[8] S. S. Eikenberry,[31] M. Eisenmann,[36] R. A. Eisenstein,[54] A. Ejlli,[113] L. Errico,[87,5] R. C. Essick,[102] H. Estelles,[110] D. Estevez,[36] Z. B. Etienne,[131] T. Etzel,[1] M. Evans,[54] T. M. Evans,[7] B. E. Ewing,[132] V. Fafone,[93,34,17] S. Fairhurst,[113] X. Fan,[91] S. Farinon,[64] B. Farr,[78] W. M. Farr,[100,101] E. J. Fauchon-Jones,[113] M. Favata,[38] M. Fays,[121] M. Fazio,[133] J. Feicht,[1] M. M. Fejer,[56] F. Feng,[27] E. Fenyvesi,[55,134] D. L. Ferguson,[85] A. Fernandez-Galiana,[54] I. Ferrante,[57,21] E. C. Ferreira,[15] T. A. Ferreira,[15] F. Fidecaro,[57,21] I. Fiori,[30] D. Fiorucci,[17,18] M. Fishbach,[102] R. P. Fisher,[41] R. Fittipaldi,[135,75] M. Fitz-Axen,[46] V. Fiumara,[136,75] R. Flaminio,[36,137] E. Floden,[46] E. Flynn,[28] H. Fong,[90] J. A. Font,[95,138] P. W. F. Forsyth,[8] J.-D. Fournier,[72] S. Frasca,[127,35] F. Frasconi,[21] Z. Frei,[116] A. Freise,[13] R. Frey,[78] V. Frey,[29] P. Fritschel,[54] V. V. Frolov,[7] G. Fronzè,[139] P. Fulda,[31] M. Fyffe,[7] H. A. Gabbard,[53] B. U. Gadre,[83] S. M. Gaebel,[13] J. R. Gair,[83] S. Galaudage,[6] D. Ganapathy,[54] A. Ganguly,[19]

S. G. Gaonkar,[3] C. García-Quirós,[110] F. Garufi,[87,5] B. Gateley,[49] S. Gaudio,[37] V. Gayathri,[97] G. Gemme,[64] E. Genin,[30] A. Gennai,[21] D. George,[20] J. George,[65] L. Gergely,[140] S. Ghonge,[85] Abhirup Ghosh,[83] Archisman Ghosh,[141–143,39] S. Ghosh,[23] B. Giacomazzo,[125,126] J. A. Giaime,[2,7] K. D. Giardina,[7] D. R. Gibson,[67] C. Gier,[24] K. Gill,[33] J. Glanzer,[2] J. Gniesmer,[118] P. Godwin,[132] E. Goetz,[2,94] R. Goetz,[31] N. Gohlke,[9,10] B. Goncharov,[6] G. González,[2] A. Gopakumar,[144] S. E. Gossan,[1] M. Gosselin,[30,57,21] R. Gouaty,[36] B. Grace,[8] A. Grado,[145,5] M. Granata,[22] A. Grant,[53] S. Gras,[54] P. Grassia,[1] C. Gray,[49] R. Gray,[53] G. Greco,[70,71] A. C. Green,[31] R. Green,[113] E. M. Gretarsson,[37] H. L. Griggs,[85] G. Grignani,[42,43] A. Grimaldi,[125,126] S. J. Grimm,[17,18] H. Grote,[113] S. Grunewald,[83] P. Gruning,[29] G. M. Guidi,[70,71] A. R. Guimaraes,[2] G. Guixé,[51] H. K. Gulati,[119] Y. Guo,[39] A. Gupta,[132] Anchal Gupta,[1] P. Gupta,[39] E. K. Gustafson,[1] R. Gustafson,[146] L. Haegel,[110] O. Halim,[18,17] E. D. Hall,[54] E. Z. Hamilton,[113] G. Hammond,[53] M. Haney,[76] M. M. Hanke,[9,10] J. Hanks,[49] C. Hanna,[132] M. D. Hannam,[113] O. A. Hannuksela,[103] T. J. Hansen,[37] J. Hanson,[7] T. Harder,[72] T. Hardwick,[2] K. Haris,[19] J. Harms,[17,18] G. M. Harry,[147] I. W. Harry,[130] R. K. Hasskew,[7] C.-J. Haster,[54] K. Haughian,[53] F. J. Hayes,[53] J. Healy,[68] A. Heidmann,[79] M. C. Heintze,[7] J. Heinze,[9,10] H. Heitmann,[72] F. Hellman,[148] P. Hello,[29] G. Hemming,[30] M. Hendry,[53] I. S. Heng,[53] E. Hennes,[39] J. Hennig,[9,10] M. Heurs,[9,10] S. Hild,[149,53] T. Hinderer,[143,39,141] S. Y. Hoback,[28,147] S. Hochheim,[9,10] E. Hofgard,[56] D. Hofman,[22] A. M. Holgado,[20] N. A. Holland,[8] K. Holt,[7] D. E. Holz,[102] P. Hopkins,[113] C. Horst,[23] J. Hough,[53] E. J. Howell,[73] C. G. Hoy,[113] Y. Huang,[54] M. T. Hübner,[6] E. A. Huerta,[20] D. Huet,[29] B. Hughey,[37] V. Hui,[36] S. Husa,[110] S. H. Huttner,[53] R. Huxford,[132] T. Huynh-Dinh,[7] B. Idzkowski,[81] A. Iess,[93,34] H. Inchauspe,[31] C. Ingram,[63] G. Intini,[127,35] J.-M. Isac,[79] M. Isi,[54] B. R. Iyer,[19] T. Jacqmin,[79] S. J. Jadhav,[150] S. P. Jadhav,[3] A. L. James,[113] K. Jani,[85] N. N. Janthalur,[150] P. Jaranowski,[151] D. Jariwala,[31] R. Jaume,[110] A. C. Jenkins,[152] J. Jiang,[31] G. R. Johns,[41] N. K. Johnson-McDaniel,[12] A. W. Jones,[13] D. I. Jones,[153] J. D. Jones,[49] P. Jones,[13] R. Jones,[53] R. J. G. Jonker,[39] L. Ju,[73] J. Junker,[9,10] C. V. Kalaghatgi,[113] V. Kalogera,[14] B. Kamai,[1] S. Kandhasamy,[3] G. Kang,[40] J. B. Kanner,[1] S. J. Kapadia,[19] S. Karki,[78] R. Kashyap,[19] M. Kasprzack,[1] W. Kastaun,[9,10] S. Katsanevas,[30] E. Katsavounidis,[54] W. Katzman,[7] S. Kaufer,[10] K. Kawabe,[49] F. Kéfélian,[72] D. Keitel,[130] A. Keivani,[33] R. Kennedy,[121] J. S. Key,[154] S. Khadka,[56] F. Y. Khalili,[66] I. Khan,[17,34] S. Khan,[9,10] Z. A. Khan,[91] E. A. Khazanov,[155] N. Khetan,[17,18] M. Khursheed,[65] N. Kijbunchoo,[8] Chunglee Kim,[156] G. J. Kim,[85] J. C. Kim,[157] K. Kim,[103] W. Kim,[63] W. S. Kim,[158] Y.-M. Kim,[159] C. Kimball,[14] P. J. King,[49] M. Kinley-Hanlon,[53] R. Kirchhoff,[9,10] J. S. Kissel,[49] L. Kleybolte,[118] S. Klimenko,[31] T. D. Knowles,[131] E. Knyazev,[54] P. Koch,[9,10] S. M. Koehlenbeck,[9,10] G. Koekoek,[39,149] S. Koley,[39] V. Kondrashov,[1] A. Kontos,[160] N. Koper,[9,10] M. Korobko,[118] W. Z. Korth,[1] M. Kovalam,[73] D. B. Kozak,[1] V. Kringel,[9,10] N. V. Krishnendu,[32] A. Królak,[161,162] N. Krupinski,[23] G. Kuehn,[9,10] A. Kumar,[150] P. Kumar,[163] Rahul Kumar,[49] Rakesh Kumar,[119] S. Kumar,[19] L. Kuo,[98] A. Kutynia,[161] B. D. Lackey,[83] D. Laghi,[57,21] E. Lalande,[164] T. L. Lam,[103] A. Lamberts,[72,165] M. Landry,[49] B. B. Lane,[54] R. N. Lang,[166] J. Lange,[68] B. Lantz,[56] R. K. Lanza,[54] I. La Rosa,[36] A. Lartaux-Vollard,[29] P. D. Lasky,[6] M. Laxen,[7] A. Lazzarini,[1] C. Lazzaro,[60] P. Leaci,[127,35] S. Leavey,[9,10] Y. K. Lecoeuche,[49] C. H. Lee,[107] H. M. Lee,[167] H. W. Lee,[157] J. Lee,[106] K. Lee,[56] J. Lehmann,[9,10] N. Leroy,[29] N. Letendre,[36] Y. Levin,[6] A. K. Y. Li,[103] J. Li,[91] K. li,[103] T. G. F. Li,[103] X. Li,[50] F. Linde,[168,39] S. D. Linker,[117] J. N. Linley,[53] T. B. Littenberg,[169] J. Liu,[9,10] X. Liu,[23] M. Llorens-Monteagudo,[95] R. K. L. Lo,[1] A. Lockwood,[170] L. T. London,[54] A. Longo,[171,172] M. Lorenzini,[17,18] V. Loriette,[173] M. Lormand,[7] G. Losurdo,[21] J. D. Lough,[9,10] C. O. Lousto,[68] G. Lovelace,[28] H. Lück,[10,9] D. Lumaca,[93,34] A. P. Lundgren,[130] Y. Ma,[50] R. Macas,[113] S. Macfoy,[24] M. MacInnis,[54] D. M. Macleod,[113] I. A. O. MacMillan,[147] A. Macquet,[72] I. Magaña Hernandez,[23] F. Magaña-Sandoval,[31] R. M. Magee,[132] E. Majorana,[35] I. Maksimovic,[173] A. Malik,[65] N. Man,[72] V. Mandic,[46] V. Mangano,[53,127,35] G. L. Mansell,[49,54] M. Manske,[23] M. Mantovani,[30] M. Mapelli,[59,60] F. Marchesoni,[58,43,174] F. Marion,[36] S. Márka,[33] Z. Márka,[33] C. Markakis,[12] A. S. Markosyan,[56] A. Markowitz,[1] E. Maros,[1] A. Marquina,[112] S. Marsat,[27] F. Martelli,[70,71] I. W. Martin,[53] R. M. Martin,[38] V. Martinez,[86] D. V. Martynov,[13] H. Masalehdan,[118] K. Mason,[54] E. Massera,[121] A. Masserot,[36] T. J. Massinger,[54] M. Masso-Reid,[53] S. Mastrogiovanni,[27] A. Matas,[83] F. Matichard,[1,54] N. Mavalvala,[54] E. Maynard,[2] J. J. McCann,[73] R. McCarthy,[49] D. E. McClelland,[8] S. McCormick,[7] L. McCuller,[54] S. C. McGuire,[175] C. McIsaac,[130] J. McIver,[1] D. J. McManus,[8] T. McRae,[8] S. T. McWilliams,[131] D. Meacher,[23] G. D. Meadors,[6] M. Mehmet,[9,10] A. K. Mehta,[19] E. Mejuto Villa,[124,75] A. Melatos,[109] G. Mendell,[49] R. A. Mercer,[23] L. Mereni,[22] K. Merfeld,[78] E. L. Merilh,[49] J. D. Merritt,[78] M. Merzougui,[72] S. Meshkov,[1] C. Messenger,[53] C. Messick,[176] R. Metzdorff,[79] P. M. Meyers,[109] F. Meylahn,[9,10] A. Mhaske,[3] A. Miani,[125,126] H. Miao,[13] I. Michaloliakos,[31] C. Michel,[22] H. Middleton,[109] L. Milano,[87,5] A. L. Miller,[31,127,35] M. Millhouse,[109] J. C. Mills,[113] E. Milotti,[177,26] M. C. Milovich-Goff,[117] O. Minazzoli,[72,178] Y. Minenkov,[34] A. Mishkin,[31] C. Mishra,[179] T. Mistry,[121] S. Mitra,[3] V. P. Mitrofanov,[66] G. Mitselmakher,[31] R. Mittleman,[54] G. Mo,[54] K. Mogushi,[94] S. R. P. Mohapatra,[54] S. R. Mohite,[23] M. Molina-Ruiz,[148] M. Mondin,[117] M. Montani,[70,71] C. J. Moore,[13] D. Moraru,[49]

F. Morawski,[62] G. Moreno,[49] S. Morisaki,[90] B. Mours,[180] C. M. Mow-Lowry,[13] S. Mozzon,[130] F. Muciaccia,[127,35] Arunava Mukherjee,[53] D. Mukherjee,[132] S. Mukherjee,[16] Subroto Mukherjee,[119] N. Mukund,[9,10] A. Mullavey,[7] J. Munch,[63] E. A. Muñiz,[44] P. G. Murray,[53] A. Nagar,[96,139,181] I. Nardecchia,[93,34] L. Naticchioni,[127,35] R. K. Nayak,[182] B. F. Neil,[73] J. Neilson,[124,75] G. Nelemans,[183,39] T. J. N. Nelson,[7] M. Nery,[9,10] A. Neunzert,[146] K. Y. Ng,[54] S. Ng,[63] C. Nguyen,[27] P. Nguyen,[78] D. Nichols,[143,39] S. A. Nichols,[2] S. Nissanke,[143,39] A. Nitz,[9,10] F. Nocera,[30] M. Noh,[54] C. North,[113] D. Nothard,[184] L. K. Nuttall,[130] J. Oberling,[49] B. D. O'Brien,[31] G. Oganesyan,[17,18] G. H. Ogin,[185] J. J. Oh,[158] S. H. Oh,[158] F. Ohme,[9,10] H. Ohta,[90] M. A. Okada,[15] M. Oliver,[110] C. Olivetto,[30] P. Oppermann,[9,10] Richard J. Oram,[7] B. O'Reilly,[7] R. G. Ormiston,[46] L. F. Ortega,[31] R. O'Shaughnessy,[68] S. Ossokine,[83] C. Osthelder,[1] D. J. Ottaway,[63] H. Overmier,[7] B. J. Owen,[92] A. E. Pace,[132] G. Pagano,[57,21] M. A. Page,[73] G. Pagliaroli,[17,18] A. Pai,[97] S. A. Pai,[65] J. R. Palamos,[78] O. Palashov,[155] C. Palomba,[35] H. Pan,[98] P. K. Panda,[150] P. T. H. Pang,[39] C. Pankow,[14] F. Pannarale,[127,35] B. C. Pant,[65] F. Paoletti,[21] A. Paoli,[30] A. Parida,[3] W. Parker,[7,175] D. Pascucci,[53,39] A. Pasqualetti,[30] R. Passaquieti,[57,21] D. Passuello,[21] B. Patricelli,[57,21] E. Payne,[6] B. L. Pearlstone,[53] T. C. Pechsiri,[31] A. J. Pedersen,[44] M. Pedraza,[1] A. Pele,[7] S. Penn,[186] A. Perego,[125,126] C. J. Perez,[49] C. Périgois,[36] A. Perreca,[125,126] S. Perriès,[105] J. Petermann,[118] H. P. Pfeiffer,[83] M. Phelps,[9,10] K. S. Phukon,[3,168,39] O. J. Piccinni,[127,35] M. Pichot,[72] M. Piendibene,[57,21] F. Piergiovanni,[70,71] V. Pierro,[124,75] G. Pillant,[30] L. Pinard,[22] I. M. Pinto,[124,75,96] K. Piotrzkowski,[80] M. Pirello,[49] M. Pitkin,[187] W. Plastino,[171,172] R. Poggiani,[57,21] D. Y. T. Pong,[103] S. Ponrathnam,[3] P. Popolizio,[30] E. K. Porter,[27] J. Powell,[188] A. K. Prajapati,[119] K. Prasai,[56] R. Prasanna,[150] G. Pratten,[13] T. Prestegard,[23] M. Principe,[124,96,75] G. A. Prodi,[125,126] L. Prokhorov,[13] M. Punturo,[43] P. Puppo,[35] M. Pürrer,[83] H. Qi,[113] V. Quetschke,[16] P. J. Quinonez,[37] F. J. Raab,[49] G. Raaijmakers,[143,39] H. Radkins,[49] N. Radulesco,[72] P. Raffai,[116] H. Rafferty,[189] S. Raja,[65] C. Rajan,[65] B. Rajbhandari,[92] M. Rakhmanov,[16] K. E. Ramirez,[16] A. Ramos-Buades,[110] Javed Rana,[3] K. Rao,[14] P. Rapagnani,[127,35] V. Raymond,[113] M. Razzano,[57,21] J. Read,[28] T. Regimbau,[36] L. Rei,[64] S. Reid,[24] D. H. Reitze,[1,31] P. Rettegno,[139,190] F. Ricci,[127,35] C. J. Richardson,[37] J. W. Richardson,[1] P. M. Ricker,[20] G. Riemenschneider,[190,139] K. Riles,[146] M. Rizzo,[14] N. A. Robertson,[1,53] F. Robinet,[29] A. Rocchi,[34] R. D. Rodriguez-Soto,[37] L. Rolland,[36] J. G. Rollins,[1] V. J. Roma,[78] M. Romanelli,[77] R. Romano,[4,5] C. L. Romel,[49] I. M. Romero-Shaw,[6] J. H. Romie,[7] C. A. Rose,[23] D. Rose,[28] K. Rose,[184] D. Rosińska,[81] S. G. Rosofsky,[20] M. P. Ross,[170] S. Rowan,[53] S. J. Rowlinson,[13] P. K. Roy,[16] Santosh Roy,[3] Soumen Roy,[191] P. Ruggi,[30] G. Rutins,[67] K. Ryan,[49] S. Sachdev,[132] T. Sadecki,[49] M. Sakellariadou,[152] O. S. Salafia,[192,47,48] L. Salconi,[30] M. Saleem,[32] F. Salemi,[125] A. Samajdar,[39] E. J. Sanchez,[1] L. E. Sanchez,[1] N. Sanchis-Gual,[193] J. R. Sanders,[194] K. A. Santiago,[38] E. Santos,[72] N. Sarin,[6] B. Sassolas,[22] B. S. Sathyaprakash,[132,113] O. Sauter,[36] R. L. Savage,[49] V. Savant,[3] D. Sawant,[97] S. Sayah,[22] D. Schaetzl,[1] P. Schale,[78] M. Scheel,[50] J. Scheuer,[14] P. Schmidt,[13] R. Schnabel,[118] R. M. S. Schofield,[78] A. Schönbeck,[118] E. Schreiber,[9,10] B. W. Schulte,[9,10] B. F. Schutz,[113] O. Schwarm,[185] E. Schwartz,[7] J. Scott,[53] S. M. Scott,[8] E. Seidel,[20] D. Sellers,[7] A. S. Sengupta,[191] N. Sennett,[83] D. Sentenac,[30] V. Sequino,[64] A. Sergeev,[155] Y. Setyawati,[9,10] D. A. Shaddock,[8] T. Shaffer,[49] S. Sharifi,[2] M. S. Shahriar,[14] A. Sharma,[17,18] P. Sharma,[65] P. Shawhan,[84] H. Shen,[20] M. Shikauchi,[90] R. Shink,[164] D. H. Shoemaker,[54] D. M. Shoemaker,[85] K. Shukla,[148] S. ShyamSundar,[65] K. Siellez,[85] M. Sieniawska,[62] D. Sigg,[49] L. P. Singer,[88] D. Singh,[132] N. Singh,[81] A. Singha,[53] A. Singhal,[17,35] A. M. Sintes,[110] V. Sipala,[128,129] V. Skliris,[113] B. J. J. Slagmolen,[8] T. J. Slaven-Blair,[73] J. Smetana,[13] J. R. Smith,[28] R. J. E. Smith,[6] S. Somala,[195] E. J. Son,[158] S. Soni,[2] B. Sorazu,[53] V. Sordini,[105] F. Sorrentino,[64] T. Souradeep,[3] E. Sowell,[92] A. P. Spencer,[53] M. Spera,[59,60] A. K. Srivastava,[119] V. Srivastava,[44] K. Staats,[14] C. Stachie,[72] M. Standke,[9,10] D. A. Steer,[27] M. Steinke,[9,10] J. Steinlechner,[118,53] S. Steinlechner,[118] D. Steinmeyer,[9,10] S. Stevenson,[188] D. Stocks,[56] D. J. Stops,[13] M. Stover,[184] K. A. Strain,[53] G. Stratta,[196,71] A. Strunk,[49] R. Sturani,[197] A. L. Stuver,[198] S. Sudhagar,[3] V. Sudhir,[54] T. Z. Summerscales,[199] L. Sun,[1] S. Sunil,[119] A. Sur,[62] J. Suresh,[90] P. J. Sutton,[113] B. L. Swinkels,[39] M. J. Szczepańczyk,[31] M. Tacca,[39] S. C. Tait,[53] C. Talbot,[6] A. J. Tanasijczuk,[80] D. B. Tanner,[31] D. Tao,[1] M. Tápai,[140] A. Tapia,[28] E. N. Tapia San Martin,[39] J. D. Tasson,[200] R. Taylor,[1] R. Tenorio,[110] L. Terkowski,[118] M. P. Thirugnanasambandam,[3] M. Thomas,[7] P. Thomas,[49] J. E. Thompson,[113] S. R. Thondapu,[65] K. A. Thorne,[7] E. Thrane,[6] C. L. Tinsman,[6] T. R. Saravanan,[3] Shubhanshu Tiwari,[76,125,126] S. Tiwari,[144] V. Tiwari,[113] K. Toland,[53] M. Tonelli,[57,21] Z. Tornasi,[53] A. Torres-Forné,[83] C. I. Torrie,[1] I. Tosta e Melo,[128,129] D. Töyrä,[8] F. Travasso,[58,43] G. Traylor,[7] M. C. Tringali,[81] A. Tripathee,[146] A. Trovato,[27] R. J. Trudeau,[1] K. W. Tsang,[39] M. Tse,[54] R. Tso,[50] L. Tsukada,[90] D. Tsuna,[90] T. Tsutsui,[90] M. Turconi,[72] A. S. Ubhi,[13] R. Udall,[85] K. Ueno,[90] D. Ugolini,[189] C. S. Unnikrishnan,[144] A. L. Urban,[2] S. A. Usman,[102] A. C. Utina,[53] H. Vahlbruch,[10] G. Vajente,[1] G. Valdes,[2] M. Valentini,[125,126] N. van Bakel,[39] M. van Beuzekom,[39] J. F. J. van den Brand,[82,149,39] C. Van Den Broeck,[39,201] D. C. Vander-Hyde,[44] L. van der Schaaf,[39] J. V. Van Heijningen,[73] A. A. van Veggel,[53] M. Vardaro,[168,39] V. Varma,[50] S. Vass,[1] M. Vasúth,[55] A. Vecchio,[13]

G. Vedovato,[60] J. Veitch,[53] P. J. Veitch,[63] K. Venkateswara,[170] G. Venugopalan,[1] D. Verkindt,[36] D. Veske,[33] F. Vetrano,[70,71] A. Viceré,[70,71] A. D. Viets,[202] S. Vinciguerra,[13] D. J. Vine,[67] J.-Y. Vinet,[72] S. Vitale,[54] Francisco Hernandez Vivanco,[6] T. Vo,[44] H. Vocca,[42,43] C. Vorvick,[49] S. P. Vyatchanin,[66] A. R. Wade,[8] L. E. Wade,[184] M. Wade,[184] R. Walet,[39] M. Walker,[28] G. S. Wallace,[24] L. Wallace,[1] S. Walsh,[23] J. Z. Wang,[146] S. Wang,[20] W. H. Wang,[16] R. L. Ward,[8] Z. A. Warden,[37] J. Warner,[49] M. Was,[36] J. Watchi,[111] B. Weaver,[49] L.-W. Wei,[9,10] M. Weinert,[9,10] A. J. Weinstein,[1] R. Weiss,[54] F. Wellmann,[9,10] L. Wen,[73] P. Weßels,[9,10] J. W. Westhouse,[37] K. Wette,[8] J. T. Whelan,[68] B. F. Whiting,[31] C. Whittle,[54] D. M. Wilken,[9,10] D. Williams,[53] J. L. Willis,[1] B. Willke,[10,9] W. Winkler,[9,10] C. C. Wipf,[1] H. Wittel,[9,10] G. Woan,[53] J. Woehler,[9,10] J. K. Wofford,[68] I. C. F. Wong,[103] J. L. Wright,[53] D. S. Wu,[9,10] D. M. Wysocki,[68] L. Xiao,[1] H. Yamamoto,[1] L. Yang,[133] Y. Yang,[31] Z. Yang,[46] M. J. Yap,[8] M. Yazback,[31] D. W. Yeeles,[113] Hang Yu,[54] Haocun Yu,[54] S. H. R. Yuen,[103] A. K. Zadrożny,[16] A. Zadrożny,[161] M. Zanolin,[37] T. Zelenova,[30] J.-P. Zendri,[60] M. Zevin,[14] J. Zhang,[73] L. Zhang,[1] T. Zhang,[53] C. Zhao,[73] G. Zhao,[111] M. Zhou,[14] Z. Zhou,[14] X. J. Zhu,[6] A. B. Zimmerman,[176] M. E. Zucker,[54,1] and J. Zweizig[1]

(LIGO Scientific Collaboration and Virgo Collaboration)

[1]*LIGO, California Institute of Technology, Pasadena, California 91125, USA*
[2]*Louisiana State University, Baton Rouge, Louisiana 70803, USA*
[3]*Inter-University Centre for Astronomy and Astrophysics, Pune 411007, India*
[4]*Dipartimento di Farmacia, Università di Salerno, I-84084 Fisciano, Salerno, Italy*
[5]*INFN, Sezione di Napoli, Complesso Universitario di Monte S.Angelo, I-80126 Napoli, Italy*
[6]*OzGrav, School of Physics & Astronomy, Monash University, Clayton 3800, Victoria, Australia*
[7]*LIGO Livingston Observatory, Livingston, Louisiana 70754, USA*
[8]*OzGrav, Australian National University, Canberra, Australian Capital Territory 0200, Australia*
[9]*Max Planck Institute for Gravitational Physics (Albert Einstein Institute), D-30167 Hannover, Germany*
[10]*Leibniz Universität Hannover, D-30167 Hannover, Germany*
[11]*Theoretisch-Physikalisches Institut, Friedrich-Schiller-Universität Jena, D-07743 Jena, Germany*
[12]*University of Cambridge, Cambridge CB2 1TN, United Kingdom*
[13]*University of Birmingham, Birmingham B15 2TT, United Kingdom*
[14]*Center for Interdisciplinary Exploration & Research in Astrophysics (CIERA), Northwestern University, Evanston, Illinois 60208, USA*
[15]*Instituto Nacional de Pesquisas Espaciais, 12227-010 São José dos Campos, São Paulo, Brazil*
[16]*The University of Texas Rio Grande Valley, Brownsville, Texas 78520, USA*
[17]*Gran Sasso Science Institute (GSSI), I-67100 L'Aquila, Italy*
[18]*INFN, Laboratori Nazionali del Gran Sasso, I-67100 Assergi, Italy*
[19]*International Centre for Theoretical Sciences, Tata Institute of Fundamental Research, Bengaluru 560089, India*
[20]*NCSA, University of Illinois at Urbana-Champaign, Urbana, Illinois 61801, USA*
[21]*INFN, Sezione di Pisa, I-56127 Pisa, Italy*
[22]*Laboratoire des Matériaux Avancés (LMA), IP2I—UMR 5822, CNRS, Université de Lyon, F-69622 Villeurbanne, France*
[23]*University of Wisconsin-Milwaukee, Milwaukee, Wisconsin 53201, USA*
[24]*SUPA, University of Strathclyde, Glasgow G1 1XQ, United Kingdom*
[25]*Dipartimento di Matematica e Informatica, Università di Udine, I-33100 Udine, Italy*
[26]*INFN, Sezione di Trieste, I-34127 Trieste, Italy*
[27]*APC, AstroParticule et Cosmologie, Université Paris Diderot, CNRS/IN2P3, CEA/Irfu, Observatoire de Paris, Sorbonne Paris Cité, F-75205 Paris Cedex 13, France*
[28]*California State University Fullerton, Fullerton, California 92831, USA*
[29]*LAL, Université Paris-Sud, CNRS/IN2P3, Université Paris-Saclay, F-91898 Orsay, France*
[30]*European Gravitational Observatory (EGO), I-56021 Cascina, Pisa, Italy*
[31]*University of Florida, Gainesville, Florida 32611, USA*
[32]*Chennai Mathematical Institute, Chennai 603103, India*
[33]*Columbia University, New York, New York 10027, USA*
[34]*INFN, Sezione di Roma Tor Vergata, I-00133 Roma, Italy*
[35]*INFN, Sezione di Roma, I-00185 Roma, Italy*
[36]*Laboratoire d'Annecy de Physique des Particules (LAPP), Université Grenoble Alpes, Université Savoie Mont Blanc, CNRS/IN2P3, F-74941 Annecy, France*
[37]*Embry-Riddle Aeronautical University, Prescott, Arizona 86301, USA*
[38]*Montclair State University, Montclair, New Jersey 07043, USA*
[39]*Nikhef, Science Park 105, 1098 XG Amsterdam, Netherlands*

[40]Korea Institute of Science and Technology Information, Daejeon 34141, South Korea
[41]Christopher Newport University, Newport News, Virginia 23606, USA
[42]Università di Perugia, I-06123 Perugia, Italy
[43]INFN, Sezione di Perugia, I-06123 Perugia, Italy
[44]Syracuse University, Syracuse, New York 13244, USA
[45]Université de Liège, B-4000 Liège, Belgium
[46]University of Minnesota, Minneapolis, Minnesota 55455, USA
[47]Università degli Studi di Milano-Bicocca, I-20126 Milano, Italy
[48]INFN, Sezione di Milano-Bicocca, I-20126 Milano, Italy
[49]LIGO Hanford Observatory, Richland, Washington 99352, USA
[50]Caltech CaRT, Pasadena, California 91125, USA
[51]Departament de Física Quàntica i Astrofísica, Institut de Ciències del Cosmos (ICCUB), Universitat de Barcelona (IEEC-UB), E-08028 Barcelona, Spain
[52]Dipartimento di Medicina, Chirurgia e Odontoiatria "Scuola Medica Salernitana," Università di Salerno, I-84081 Baronissi, Salerno, Italy
[53]SUPA, University of Glasgow, Glasgow G12 8QQ, United Kingdom
[54]LIGO, Massachusetts Institute of Technology, Cambridge, Massachusetts 02139, USA
[55]Wigner RCP, RMKI, H-1121 Budapest, Konkoly Thege Miklós út 29-33, Hungary
[56]Stanford University, Stanford, California 94305, USA
[57]Università di Pisa, I-56127 Pisa, Italy
[58]Università di Camerino, Dipartimento di Fisica, I-62032 Camerino, Italy
[59]Università di Padova, Dipartimento di Fisica e Astronomia, I-35131 Padova, Italy
[60]INFN, Sezione di Padova, I-35131 Padova, Italy
[61]Montana State University, Bozeman, Montana 59717, USA
[62]Nicolaus Copernicus Astronomical Center, Polish Academy of Sciences, 00-716, Warsaw, Poland
[63]OzGrav, University of Adelaide, Adelaide, South Australia 5005, Australia
[64]INFN, Sezione di Genova, I-16146 Genova, Italy
[65]RRCAT, Indore, Madhya Pradesh 452013, India
[66]Faculty of Physics, Lomonosov Moscow State University, Moscow 119991, Russia
[67]SUPA, University of the West of Scotland, Paisley PA1 2BE, United Kingdom
[68]Rochester Institute of Technology, Rochester, New York 14623, USA
[69]Bar-Ilan University, Ramat Gan 5290002, Israel
[70]Università degli Studi di Urbino "Carlo Bo," I-61029 Urbino, Italy
[71]INFN, Sezione di Firenze, I-50019 Sesto Fiorentino, Firenze, Italy
[72]Artemis, Université Côte d'Azur, Observatoire Côte d'Azur, CNRS, CS 34229, F-06304 Nice Cedex 4, France
[73]OzGrav, University of Western Australia, Crawley, Western Australia 6009, Australia
[74]Dipartimento di Fisica "E.R. Caianiello," Università di Salerno, I-84084 Fisciano, Salerno, Italy
[75]INFN, Sezione di Napoli, Gruppo Collegato di Salerno, Complesso Universitario di Monte S. Angelo, I-80126 Napoli, Italy
[76]Physik-Institut, University of Zurich, Winterthurerstrasse 190, 8057 Zurich, Switzerland
[77]Université Rennes, CNRS, Institut FOTON—UMR6082, F-3500 Rennes, France
[78]University of Oregon, Eugene, Oregon 97403, USA
[79]Laboratoire Kastler Brossel, Sorbonne Université, CNRS, ENS-Université PSL, Collège de France, F-75005 Paris, France
[80]Université catholique de Louvain, B-1348 Louvain-la-Neuve, Belgium
[81]Astronomical Observatory Warsaw University, 00-478 Warsaw, Poland
[82]VU University Amsterdam, 1081 HV Amsterdam, Netherlands
[83]Max Planck Institute for Gravitational Physics (Albert Einstein Institute), D-14476 Potsdam-Golm, Germany
[84]University of Maryland, College Park, Maryland 20742, USA
[85]School of Physics, Georgia Institute of Technology, Atlanta, Georgia 30332, USA
[86]Université de Lyon, Université Claude Bernard Lyon 1, CNRS, Institut Lumière Matière, F-69622 Villeurbanne, France
[87]Università di Napoli "Federico II," Complesso Universitario di Monte S.Angelo, I-80126 Napoli, Italy
[88]NASA Goddard Space Flight Center, Greenbelt, Maryland 20771, USA
[89]Dipartimento di Fisica, Università degli Studi di Genova, I-16146 Genova, Italy
[90]RESCEU, University of Tokyo, Tokyo, 113-0033, Japan
[91]Tsinghua University, Beijing 100084, China
[92]Texas Tech University, Lubbock, Texas 79409, USA
[93]Università di Roma Tor Vergata, I-00133 Roma, Italy
[94]Missouri University of Science and Technology, Rolla, Missouri 65409, USA
[95]Departamento de Astronomía y Astrofísica, Universitat de València, E-46100 Burjassot, València, Spain
[96]Museo Storico della Fisica e Centro Studi e Ricerche "Enrico Fermi," I-00184 Roma, Italy
[97]Indian Institute of Technology Bombay, Powai, Mumbai 400 076, India

[98]National Tsing Hua University, Hsinchu City, 30013 Taiwan, Republic of China
[99]Charles Sturt University, Wagga Wagga, New South Wales 2678, Australia
[100]Physics and Astronomy Department, Stony Brook University, Stony Brook, New York 11794, USA
[101]Center for Computational Astrophysics, Flatiron Institute, 162 5th Avenue, New York, New York 10010, USA
[102]University of Chicago, Chicago, Illinois 60637, USA
[103]The Chinese University of Hong Kong, Shatin, NT, Hong Kong
[104]Dipartimento di Ingegneria Industriale (DIIN), Università di Salerno, I-84084 Fisciano, Salerno, Italy
[105]Institut de Physique des 2 Infinis de Lyon (IP2I)—UMR 5822, Université de Lyon, Université Claude Bernard, CNRS, F-69622 Villeurbanne, France
[106]Seoul National University, Seoul 08826, South Korea
[107]Pusan National University, Busan 46241, South Korea
[108]INAF, Osservatorio Astronomico di Padova, I-35122 Padova, Italy
[109]OzGrav, University of Melbourne, Parkville, Victoria 3010, Australia
[110]Universitat de les Illes Balears, IAC3—IEEC, E-07122 Palma de Mallorca, Spain
[111]Université Libre de Bruxelles, Brussels 1050, Belgium
[112]Departamento de Matemáticas, Universitat de València, E-46100 Burjassot, València, Spain
[113]Cardiff University, Cardiff CF24 3AA, United Kingdom
[114]University of Rhode Island, Kingston, Rhode Island 02881, USA
[115]Bellevue College, Bellevue, Washington 98007, USA
[116]MTA-ELTE Astrophysics Research Group, Institute of Physics, Eötvös University, Budapest 1117, Hungary
[117]California State University, Los Angeles, 5151 State University Drive, Los Angeles, California 90032, USA
[118]Universität Hamburg, D-22761 Hamburg, Germany
[119]Institute for Plasma Research, Bhat, Gandhinagar 382428, India
[120]IGFAE, Campus Sur, Universidade de Santiago de Compostela, 15782 Spain
[121]The University of Sheffield, Sheffield S10 2TN, United Kingdom
[122]Dipartimento di Scienze Matematiche, Fisiche e Informatiche, Università di Parma, I-43124 Parma, Italy
[123]INFN, Sezione di Milano Bicocca, Gruppo Collegato di Parma, I-43124 Parma, Italy
[124]Dipartimento di Ingegneria, Università del Sannio, I-82100 Benevento, Italy
[125]Università di Trento, Dipartimento di Fisica, I-38123 Povo, Trento, Italy
[126]INFN, Trento Institute for Fundamental Physics and Applications, I-38123 Povo, Trento, Italy
[127]Università di Roma "La Sapienza," I-00185 Roma, Italy
[128]Università degli Studi di Sassari, I-07100 Sassari, Italy
[129]INFN, Laboratori Nazionali del Sud, I-95125 Catania, Italy
[130]University of Portsmouth, Portsmouth, PO1 3FX, United Kingdom
[131]West Virginia University, Morgantown, West Virginia 26506, USA
[132]The Pennsylvania State University, University Park, Pennsylvania 16802, USA
[133]Colorado State University, Fort Collins, Colorado 80523, USA
[134]Institute for Nuclear Research (Atomki), Hungarian Academy of Sciences, Bem tér 18/c, H-4026 Debrecen, Hungary
[135]CNR-SPIN, c/o Università di Salerno, I-84084 Fisciano, Salerno, Italy
[136]Scuola di Ingegneria, Università della Basilicata, I-85100 Potenza, Italy
[137]National Astronomical Observatory of Japan, 2-21-1 Osawa, Mitaka, Tokyo 181-8588, Japan
[138]Observatori Astronòmic, Universitat de València, E-46980 Paterna, València, Spain
[139]INFN Sezione di Torino, I-10125 Torino, Italy
[140]University of Szeged, Dóm tér 9, Szeged 6720, Hungary
[141]Delta Institute for Theoretical Physics, Science Park 904, 1090 GL Amsterdam, Netherlands
[142]Lorentz Institute, Leiden University, PO Box 9506, Leiden 2300 RA, Netherlands
[143]GRAPPA, Anton Pannekoek Institute for Astronomy and Institute for High-Energy Physics, University of Amsterdam, Science Park 904, 1098 XH Amsterdam, Netherlands
[144]Tata Institute of Fundamental Research, Mumbai 400005, India
[145]INAF, Osservatorio Astronomico di Capodimonte, I-80131 Napoli, Italy
[146]University of Michigan, Ann Arbor, Michigan 48109, USA
[147]American University, Washington, DC 20016, USA
[148]University of California, Berkeley, California 94720, USA
[149]Maastricht University, P.O. Box 616, 6200 MD Maastricht, Netherlands
[150]Directorate of Construction, Services & Estate Management, Mumbai 400094 India
[151]University of Białystok, 15-424 Białystok, Poland
[152]King's College London, University of London, London WC2R 2LS, United Kingdom
[153]University of Southampton, Southampton SO17 1BJ, United Kingdom
[154]University of Washington Bothell, Bothell, Washington 98011, USA
[155]Institute of Applied Physics, Nizhny Novgorod, 603950, Russia

[156]*Ewha Womans University, Seoul 03760, South Korea*
[157]*Inje University Gimhae, South Gyeongsang 50834, South Korea*
[158]*National Institute for Mathematical Sciences, Daejeon 34047, South Korea*
[159]*Ulsan National Institute of Science and Technology, Ulsan 44919, South Korea*
[160]*Bard College, 30 Campus Rd, Annandale-On-Hudson, New York 12504, USA*
[161]*NCBJ, 05-400 Świerk-Otwock, Poland*
[162]*Institute of Mathematics, Polish Academy of Sciences, 00656 Warsaw, Poland*
[163]*Cornell University, Ithaca, New York 14850, USA*
[164]*Université de Montréal/Polytechnique, Montreal, Quebec H3T 1J4, Canada*
[165]*Lagrange, Université Côte d'Azur, Observatoire Côte d'Azur, CNRS, CS 34229, F-06304 Nice Cedex 4, France*
[166]*Hillsdale College, Hillsdale, Michigan 49242, USA*
[167]*Korea Astronomy and Space Science Institute, Daejeon 34055, South Korea*
[168]*Institute for High-Energy Physics, University of Amsterdam, Science Park 904, 1098 XH Amsterdam, Netherlands*
[169]*NASA Marshall Space Flight Center, Huntsville, Alabama 35811, USA*
[170]*University of Washington, Seattle, Washington 98195, USA*
[171]*Dipartimento di Matematica e Fisica, Università degli Studi Roma Tre, I-00146 Roma, Italy*
[172]*INFN, Sezione di Roma Tre, I-00146 Roma, Italy*
[173]*ESPCI, CNRS, F-75005 Paris, France*
[174]*Center for Phononics and Thermal Energy Science, School of Physics Science and Engineering, Tongji University, 200092 Shanghai, People's Republic of China*
[175]*Southern University and A&M College, Baton Rouge, Louisiana 70813, USA*
[176]*Department of Physics, University of Texas, Austin, Texas 78712, USA*
[177]*Dipartimento di Fisica, Università di Trieste, I-34127 Trieste, Italy*
[178]*Centre Scientifique de Monaco, 8 quai Antoine Ier, MC-98000, Monaco*
[179]*Indian Institute of Technology Madras, Chennai 600036, India*
[180]*Université de Strasbourg, CNRS, IPHC UMR 7178, F-67000 Strasbourg, France*
[181]*Institut des Hautes Etudes Scientifiques, F-91440 Bures-sur-Yvette, France*
[182]*IISER-Kolkata, Mohanpur, West Bengal 741252, India*
[183]*Department of Astrophysics/IMAPP, Radboud University Nijmegen, P.O. Box 9010, 6500 GL Nijmegen, Netherlands*
[184]*Kenyon College, Gambier, Ohio 43022, USA*
[185]*Whitman College, 345 Boyer Avenue, Walla Walla, Washington 99362 USA*
[186]*Hobart and William Smith Colleges, Geneva, New York 14456, USA*
[187]*Department of Physics, Lancaster University, Lancaster, LA1 4YB, United Kingdom*
[188]*OzGrav, Swinburne University of Technology, Hawthorn VIC 3122, Australia*
[189]*Trinity University, San Antonio, Texas 78212, USA*
[190]*Dipartimento di Fisica, Università degli Studi di Torino, I-10125 Torino, Italy*
[191]*Indian Institute of Technology, Gandhinagar Ahmedabad Gujarat 382424, India*
[192]*INAF, Osservatorio Astronomico di Brera sede di Merate, I-23807 Merate, Lecco, Italy*
[193]*Centro de Astrofísica e Gravitação (CENTRA), Departamento de Física, Instituto Superior Técnico, Universidade de Lisboa, 1049-001 Lisboa, Portugal*
[194]*Marquette University, 11420 West Clybourn Street, Milwaukee, Wisconsin 53233, USA*
[195]*Indian Institute of Technology Hyderabad, Sangareddy, Khandi, Telangana 502285, India*
[196]*INAF, Osservatorio di Astrofisica e Scienza dello Spazio, I-40129 Bologna, Italy*
[197]*International Institute of Physics, Universidade Federal do Rio Grande do Norte, Natal RN 59078-970, Brazil*
[198]*Villanova University, 800 Lancaster Avenue, Villanova, Pennsylvania 19085, USA*
[199]*Andrews University, Berrien Springs, Michigan 49104, USA*
[200]*Carleton College, Northfield, Minnesota 55057, USA*
[201]*Department of Physics, Utrecht University, 3584CC Utrecht, Netherlands*
[202]*Concordia University Wisconsin, 2800 North Lake Shore Drive, Mequon, Wisconsin 53097, USA*